\newif\iffigs\figstrue
    \documentstyle[12pt]{article}
\iffigs
   \input{epsf}
\else
\fi
\newcommand{\eq}{\begin{equation}}
\newcommand{\eqa}{\begin{eqnarray}}
\newcommand{\en}{\end{equation}}
\newcommand{\ena}{\end{eqnarray}}
\newcommand{\enn}{\nonumber \end{equation}}

\newcommand{\eqn}[1]{(\ref{#1})}

\newsavebox{\uuunit}
\sbox{\uuunit}
    {\setlength{\unitlength}{0.825em}
     \begin{picture}(0.6,0.7)
        \thinlines
        \put(0,0){\line(1,0){0.5}}
        \put(0.15,0){\line(0,1){0.7}}
        \put(0.35,0){\line(0,1){0.8}}
       \multiput(0.3,0.8)(-0.04,-0.02){12}{\rule{0.5pt}{0.5pt}}
     \end {picture}}

\def\sk{\vskip .4cm}

\def\IP{\relax{\rm I\kern-.18em P}}

\def\muun{\underline \mu}
\def\mun{\underline m}
\def\nuun{\underline \nu}
\def\nun{\underline n}
\def\buun{\underline \bullet}
\def\Eb{{\bf E}}
\def\bu{\bullet}
\def\we{\wedge}
\begin{document}
%
\font\cmss=cmss10 \font\cmsss=cmss10 at 7pt
\def\twomat#1#2#3#4{\left(\matrix{#1 & #2 \cr #3 & #4}\right)}
\def\inbar{\vrule height1.5ex width.4pt depth0pt}
\def\IC{\relax\,\hbox{$\inbar\kern-.3em{\rm C}$}}
\def\IG{\relax\,\hbox{$\inbar\kern-.3em{\rm G}$}}
\def\IB{\relax{\rm I\kern-.18em B}}
\def\ID{\relax{\rm I\kern-.18em D}}
\def\IL{\relax{\rm I\kern-.18em L}}
\def\IF{\relax{\rm I\kern-.18em F}}
\def\IH{\relax{\rm I\kern-.18em H}}
\def\II{\relax{\rm I\kern-.17em I}}
\def\IN{\relax{\rm I\kern-.18em N}}
\def\IP{\relax{\rm I\kern-.18em P}}
\def\IQ{\relax\,\hbox{$\inbar\kern-.3em{\rm Q}$}}
\def\bfzero{\relax\,\hbox{$\inbar\kern-.3em{\rm 0}$}}
\def\IK{\relax{\rm I\kern-.18em K}}
\def\IG{\relax\,\hbox{$\inbar\kern-.3em{\rm G}$}}
 \font\cmss=cmss10 \font\cmsss=cmss10 at 7pt
\def\IR{\relax{\rm I\kern-.18em R}}
\def\ZZ{\relax\ifmmode\mathchoice
{\hbox{\cmss Z\kern-.4em Z}}{\hbox{\cmss Z\kern-.4em Z}}
{\lower.9pt\hbox{\cmsss Z\kern-.4em Z}}
{\lower1.2pt\hbox{\cmsss Z\kern-.4em Z}}\else{\cmss Z\kern-.4em
Z}\fi}
\def\bfone{\relax{\rm 1\kern-.35em 1}}
\def\dop{{\rm d}\hskip -1pt}
\def\real{{\rm Re}\hskip 1pt}
\def\trace{{\rm Tr}\hskip 1pt}
\def\ii{{\rm i}}
\def\diag{{\rm diag}}
\def\sch#1#2{\{#1;#2\}}
\def\sk{\vskip .4cm}
\def\noi{\noindent}
\def\om{\omega}
\def\Om{\Omega}
\def\al{\alpha}
\def\la{\lambda}
\def\be{\beta}
\def\ga{\gamma}
\def\Ga{\Gamma}
\def\de{\delta}
\def\epsi{\varepsilon}
\def\we{\wedge}
\def\part{\partial}
\def\bu{\bullet}
\def\ci{\circ}
\def\square{{\,\lower0.9pt\vbox{\hrule \hbox{\vrule height 0.2 cm
\hskip 0.2 cm \vrule height 0.2 cm}\hrule}\,}}
\def\muun{\underline \mu}
\def\mun{\underline m}
\def\nuun{\underline \nu}
\def\nun{\underline n}
\def\buun{\underline \bullet}
\def\Rb{{\bf R}}
\def\Eb{{\bf E}}
\def\gb{{\bf g}}
\def\dt{{\tilde d}}
\def\Dt{{\tilde D}}
\def\Dcal{{\cal D}}
\def\R#1#2{ R^{#1}_{~~~#2} }
\def\ome#1#2{\om^{#1}_{~#2}}
\def\Rf#1#2{ R^{\underline #1}_{~~~{\underline #2}} }
\def\Rfu#1#2{ R^{{\underline #1}{\underline #2}} }
\def\Rfd#1#2{ R_{{\underline #1}{\underline #2}} }
\def\Rfb#1#2{ {\bf R}^{\underline #1}_{~~~{\underline #2}} }
\def\omef#1#2{\om^{\underline #1}_{~{\underline #2}}}
\def\omefb#1#2{{ \omb}^{\underline #1}_{~{\underline #2}}}
\def\omefu#1#2{\om^{{\underline #1} {\underline #2}}}
\def\omefub#1#2{{\omb}^{{\underline #1} {\underline #2}}}
\def\Ef#1{E^{\underline #1}}
\def\Efb#1{{\bf E}^{\underline #1}}
\def\omb{\bf \mbox{\boldmath $\om$}}
\def\bfone{\relax{\rm 1\kern-.35em 1}}
\font\cmss=cmss10 \font\cmsss=cmss10 at 7pt
\def\a{\alpha} \def\b{\beta} \def\d{\delta}
\def\e{\epsilon} \def\c{\gamma}
\def\G{\Gamma} \def\l{\lambda}
\def\L{\Lambda} \def\s{\sigma}
\def\cA{{\cal A}} \def\cB{{\cal B}}
\def\cC{{\cal C}} \def\cD{{\cal D}}
\def\cF{{\cal F}} \def\cG{{\cal G}}
\def\cH{{\cal H}} \def\cI{{\cal I}}
\def\cJ{{\cal J}} \def\cK{{\cal K}}
\def\cL{{\cal L}} \def\cM{{\cal M}}
\def\cN{{\cal N}} \def\cO{{\cal O}}
\def\cP{{\cal P}} \def\cQ{{\cal Q}}
\def\cR{{\cal R}} \def\cV{{\cal V}}\def\cW{{\cal W}}
%
%
%
\def\crr{\crcr\noalign{\vskip {8.3333pt}}}
\def\tilde{\widetilde}
\def\bar{\overline}
\def\us#1{\underline{#1}}
\let\shat=\hat
\def\hat{\widehat}
\def\hyp{\vrule height 2.3pt width 2.5pt depth -1.5pt}
\def\Coeff#1#2{{#1\over #2}}
\def\Coe#1.#2.{{#1\over #2}}
\def\coeff#1#2{\relax{\textstyle {#1 \over #2}}\displaystyle}
\def\coe#1.#2.{\relax{\textstyle {#1 \over #2}}\displaystyle}
\def\half{{1 \over 2}}
\def\shalf{\relax{\textstyle {1 \over 2}}\displaystyle}
\def\dag#1{#1\!\!\!/\,\,\,}
\def\to{\rightarrow}
\def\notin{\hbox{{$\in$}\kern-.51em\hbox{/}}}
\def\shdot{\!\cdot\!}
\def\ket#1{\,\big|\,#1\,\big>\,}
\def\bra#1{\,\big<\,#1\,\big|\,}
\def\equaltop#1{\mathrel{\mathop=^{#1}}}
\def\Trbel#1{\mathop{{\rm Tr}}_{#1}}
\def\inserteq#1{\noalign{\vskip-.2truecm\hbox{#1\hfil}
\vskip-.2cm}}
\def\attac#1{\Bigl\vert
{\phantom{X}\atop{{\rm\scriptstyle #1}}\phantom{X}}}
\def\exx#1{e^{{\displaystyle #1}}}
\def\del{\partial}
\def\delbar{\bar\partial}
\def\nex#1{$N\!=\!#1$}
\def\dex#1{$d\!=\!#1$}
\def\cex#1{$c\!=\!#1$}
\def\eg{{\it e.g.}} \def\ie{{\it i.e.}}
\def\IE{\relax{{\rm I\kern-.18em E}}}
\def\cE{{\cal E}}
\def\rt{{\cR^{(3)}}}
\def\IGam{\relax{{\rm I}\kern-.18em \Gamma}}
\def\IGa{\IA}
\def\ii{{\rm i}}
\begin{titlepage}
\begin{center}
{{\LARGE   $G/H$ M--branes}\\
\vspace{6pt}
{\LARGE   and  $AdS_{p+2}$ Geometries$^*$$^{\dagger}$}
}\\
\vfill
{\large  Leonardo Castellani$^1$, Anna Ceresole$^2$,
Riccardo D'Auria $^2$,
\\
\vskip 0.2cm
Sergio Ferrara$^3$, Pietro Fr\'e$^4$
 and  Mario Trigiante$^5$   } \\
\vfill
{\small
$^1$ Dipartimento di Scienze e Tecnologie Avanzate, Universit\'a  di
Torino,
Sede di
Alessandria\\
and Istituto Nazionale di Fisica Nucleare (INFN) - Sezione di
Torino,
Italy\\
\vspace{6pt}
$^2$ Dipartimento di Fisica Politecnico di Torino, C.so Duca degli
Abruzzi,
24,
I-10129 Torino\\
and Istituto Nazionale di Fisica Nucleare (INFN) - Sezione di
Torino,
Italy\\
$^3$ CERN, Theoretical Division, CH 1211 Geneva, 23,
Switzerland\\
\vspace{6pt}
$^4$ Dipartimento di Fisica Teorica, Universit\'a di Torino, via P.
Giuria 1,
I-10125 Torino, \\
 Istituto Nazionale di Fisica Nucleare (INFN) - Sezione di Torino,
Italy \\
\vspace{6pt}
$^5$ Department of Physics, University of Wales Swansea, Singleton
Park,\\
 Swansea SA2 8PP, United Kingdom\\
\vspace{6pt}
}
\end{center}
\vfill
\begin{center}
{\bf Abstract}
\end{center}
{\small We discuss the class of
$BPS$ saturated $M$--branes that
are in one--to--one correspondence with the Freund--Rubin
compactifications of M--theory
on either $AdS_4 \, \times \, G/H$ or $AdS_7 \, \times \, G/H$,
where $G/H$ is any of  the seven (or four)
dimensional Einstein coset manifolds with Killing spinors classified  
long ago
in the
context of  Kaluza--Klein supergravity.
These $G/H$ M--branes, whose existence was previously pointed out
in the literature, are solitons that interpolate between flat space
at infinity and the old Kaluza--Klein compactifications at the
horizon. They preserve $N/2$ supersymmetries where $N$ is the number
of Killing spinors of the $AdS \, \times \, G/H$ vacuum. A crucial
ingredient in our discussion is the identification of a solvable
Lie algebra parametrization of the Lorentzian non compact coset
$SO(2,p+1)/SO(1,p+1)$ corresponding to anti--de Sitter space
$AdS_{p+2}$. The solvable coordinates are those naturally emerging
from the
near horizon limit of the $G/H$ $p$--brane and correspond to
the Bertotti--Robinson form of the anti--de Sitter metric. The
pull-back of
anti--de Sitter isometries on the $p$--brane world--volume contain,
in
particular, the recently found  {\it broken conformal  
transformations}.
}
\vspace{2mm} \vfill \hrule width 3.cm
{\footnotesize
 $^*$ Supported in part by   EEC  under TMR contract
 ERBFMRX-CT96-0045, in which A. Ceresole and R. D'Auria are
associated
 to Torino University and S. Ferrara  is associated to Frascati and
by
 DOE grant DE-FG03-91ER40662.
\par
$^{\dagger}$ Supported in part by EEC  under TMR contract
 ERBFMRXCT960012, in which M. Trigiante is associated to Swansea
University}
\end{titlepage}
\section{Introduction}
\label{intro}
Since the second string revolution \cite{secrevol1,secrevol2},
the five consistent $10$--dimensional superstrings have been
reinterpreted as different perturbative limits of a single
fundamental theory, named  M--theory . While
the microscopic quantum definition of
this latter is still matter of debate,  its low energy effective
lagrangian is well known and extensively studied since the end of
the
seventies: indeed it coincides with $D=11$ supergravity
\cite{sugra11a,sugra11b}. As a logical consequence of this
new deeper understanding, all aspects of $D=11$
supergravity must have bearings on string theory and admit a string
interpretation.
\par
Until a few months ago the aspect of $D=11$ supergravity whose
consequences on string theory has been  investigated most
is
given by its classical $p$--brane solutions, usually called
M--branes
 \cite{mbrastelle,mbratownsend}. In their simplest
formulation M--branes are solutions of the classical field equations
where the metric takes the following form:
\begin{equation}
ds^2_{11} \, = \, \left(1+
\frac{k}{r^{\tilde d}} \right)^{-\frac{\tilde d}{9}} \,
dx^\mu  \,   \, dx^\nu \eta_{\mu\nu} + \, \, \left(1+
\frac{k}{r^{\tilde d}} \right)^{\frac{  d}{9}} \, dX^I \,  \,
dX^J \, \delta_{IJ}\, .
\label{protyp}
\end{equation}
In eq.\eqn{protyp}
\begin{equation}
\begin{array}{ccccccc}
 d & \equiv & p+1 & ; & {\tilde d} & \equiv & 11 - d -2 \\
\end{array}
\label{ddtil}
\end{equation}
are the world--volume dimensions of the $p$--brane and of its
magnetic dual,
\begin{equation}
r \, \equiv \, \sqrt{ X^I \, X^J \, \delta_{IJ} }
\label{rdef}
\end{equation}
is the radial distance from the brane in transverse space,
$\mu=0,\ldots,d-1$ and $I,J=d,\ldots,10$.
\par
The isometry group of the $11$--dimensional metric \eqn{protyp} is:
\begin{equation}
 {\cal I}_{p-brane} \, = \, ISO(1,p) \, \otimes \, SO \left( 11-d
 \right)
 \label{isometr1}
\end{equation}
\par
There are two fundamental branes of this sort: the electric
M2--brane ($p=2$) and the magnetic M5--brane ($p=5$). In addition
one
has a variety of more complicated branes that can be interpreted as
intersections and superpositions of the fundamental ones at angles.
The basic motivation for the pre-eminence of these
M--branes in the recent studies on M--theory is their property of
being BPS states, that is of admitting a set of Killing spinors
whose existence leads to the saturation of the relevant Bogomol'nyi
bound in the mass--charge relation. Hence classical M--brane
solutions correspond to exact non--perturbative quantum states
of the string spectrum and from the string side there exist
descriptions of these states in terms of Dirichlet branes
\cite{polchidbra}
using for instance the technology of boundary states
\cite{bstastef,bstaieng,bstapaol}.
\par
One point that we would like to stress is that M--branes of type
\eqn{protyp} are asymptotically flat and have, at spatial infinity
 ($r \,  \to \, \infty$),
the same topology as the spatial infinity of $11$--dimensional
Minkowski space,
namely $S^9$.
\vskip 0.4cm
\noindent
{$\bullet$  {\bf Freund--Rubin manifolds and M--branes}}
\vskip 0.1cm
One different aspect of D=11 supergravity that was actively
investigated in the eighties
\cite{freundrub,KKduff,KKenglert,mpqr,KKwarncastel}  and that has
so far eluded being fully incorporated into M--theory is given by
the
Kaluza--Klein compactifications. In this context one has the
Freund--Rubin vacua \cite{freundrub} where the $11$--dimensional
space is either:
\begin{equation}
{\cal M}_{11} \, = \, AdS_{4} \, \times \,
\left(\frac{G}{H}\right)_7
\label{4p7}
\end{equation}
or
\begin{equation}
{\cal M}_{11} \, = \, AdS_{7} \, \times \,
\left(\frac{G}{H}\right)_4
\label{7p4}
\end{equation}
having denoted by $AdS_{D}=SO(2,D-1)/SO(1,D-1)$ anti de Sitter space
in dimension $D$ and by $\left(\frac{G}{H}\right)_n$ an
$n$--dimensional
coset manifold equipped with a $G$--invariant Einstein metric.
\par
The first class of Freund--Rubin vacua \eqn{4p7} is somehow
reminiscent of the M2--brane since such a metric solves the field
equations under the condition that the $4$--form field strength take
a constant $SO(1,3)$-invariant  vev:
\begin{equation}
F_{\mu _1\mu _2\mu _3 \mu _4} = e \, \epsilon_{\mu _1\mu _2\mu _3
\mu
_4}
\label{f4p7}
\end{equation}
on $AdS_4$. Alternatively, the second class
of Freund--Rubin vacua \eqn{4p7} is somehow
reminiscent of the M5--brane.  Indeed in this case the chosen metric
solves
the field equations if the dual magnetic $7$--form acquires
an $SO(1,6)$-invariant  vev:
\begin{equation}
^{\star}F_{\mu _1\dots \mu _7} = g \, \epsilon_{\mu _1\dots \mu _7}
\label{f7p4}
\end{equation}
on $AdS_7$.
\par
Despite this analogy,
little consideration in the context of M--theory was given to the
Freund--Rubin solutions  until the end of the
last year. \footnote{After submitting this paper, our attention was  
called
to ref. \cite{duffetal}, where
membrane solutions of $D=11$ supergravity were found, reducing to
$AdS_4 \times $ (7-dim. Einstein space) at the horizon.}
\par
Recently, in
\cite{renatoine,maldapasto} an important observation was made
that lead us to address the question  answered by the present paper.
\par
It had been known in the literature already for some years
\cite{doubling} that near the horizon ($ r \, \to \, 0$)
of an Mp--brane (defined according to \eqn{protyp}) the exact
metric becomes approximated by the metric of the following
$11$--dimensional space:
\begin{equation}
{  M}^{hor}_{p}\, =\, AdS_{p+2} \, \times \, S^{9-p}
\label{mh11}
\end{equation}
that has
\begin{equation}
{\cal I}^{hor}_p \,=\, SO(2,p+1) \, \times \, SO(10-p)
\label{Ihp}
\end{equation}
as the isometry group.
\par
It was observed in \cite{renatoine} that for $p=2$ and $p=5$ the
Lie algebra of ${\cal I}^{hor}_p$ can be identified with the bosonic
sector of a superalgebra ${\cal SC}_p$ admitting the interpretation
of conformal superalgebra on the $p$--brane world--volume.
The explicit identifications are
\begin{equation}
\begin{array}{ccccccc}
{\cal I}^{hor}_2 & = & SO(2,3) \, \times \, SO(8) & ; &
{\cal SC}_2 & = & Osp(8\vert 4) \\
\null & \null & \null & \null & \null & \null & \null \\
 {\cal I}^{hor}_5 & = & SO(2,6) \, \times \, SO(5) & ; &
{\cal SC}_5 & = & Osp(2,6\vert 4) \\
\end{array}
\label{scgroup}
\end{equation}
where $Osp(8\vert 4)$ is the real section of the complex
orthosymplectic
algebra $Osp^c(8\vert 4)$ having $SO(8) \, \times \, Sp(4,\IR)$ as
bosonic
subalgebra, while $Osp(2,6\vert 4)$ is the real section of the same
complex superalgebra having $SO(2,6) \, \times \, \left(USp(4)\sim
SO(5) \right)$ as bosonic subalgebra.
\par
In \cite{renatoine} it was shown how to realize the transformations
of ${\cal SC}_p$ as symmetries of the linearized $p$--brane
world--volume action. In \cite{maldapasto} it was instead proved
that
the non--linear Born Infeld effective action of the $p$--brane is
invariant under conformal like transformations that realize the
group
${\cal I}^{hor}_p$. Since these transformations are similar but not
identical to the standard conformal transformations the author of
\cite{maldapasto} named them {\it broken conformal transformations}.
Further very recent developments in this direction appeared
in \cite{holow,nairdaemi}.
\par
What we would like to stress in relation with these developments is
that the near horizon manifold \eqn{mh11} is just one
instance of Freund--Rubin solution, where the internal coset
manifolds
have been chosen to be:
\begin{equation}
\begin{array}{rccclcrcccl}
 \left(\frac{G}{H}\right)_7 & = & \frac{SO(8)}{SO(7)} & \equiv & S^7
& ; &
 \left(\frac{G}{H}\right)_4 & = & \frac{SO(5)}{SO(4)} & \equiv & S^4
\end{array}
\label{ghspec}
\end{equation}
and the superconformal algebras \eqn{scgroup} are nothing else but
the full supersymmetry algebras of these Freund--Rubin vacua.
\par
Therefore we see that ordinary M2 and M5--branes are correctly
interpreted as  {\it D=11 supergravity solitons} that interpolate
between two vacuum solutions of the theory, the space
\begin{equation}
M^{\infty}_p \, = \, \mbox{Minkowski}_{11}
\label{minkione11}
\end{equation}
at spatial infinity ($r \, \to \, \infty$) and the manifold
\eqn{mh11}
near the horizon ($r \, \to \, 0$), which is nothing else but one of
the possible Freund--Rubin manifolds.
\par
These solutions are $\frac{1}{2}$-BPS solitons that interpolate
between maximally symmetric geometries, i.e. flat SuperPoincar\'e
and
maximally anti de Sitter supergravity at the horizon.
\par
In this paper we consider
M--theory branes which still interpolate between flat and anti de
Sitter spaces, but with $N < N_{max}$ supersymmetries.
This allows  solutions which, at the horizon, will
look as $AdS_{p+2}$ supergravity $\times$ $ M_{D-p-2}$ where $
M_{D-p-2}$
is a manifold admitting $N$ Killing spinors, appropriate to a
$\frac{N}{2}$--BPS soliton. Solutions of this type were originally
considered in \cite{duffetal}.
\par
Recently $p$-brane theories giving $AdS_{p+2}$ horizon geometries
(for the
$p=3$ case)
with less supersymmetries have been considered via an explicit
construction of the $p+1$ world volume theory with $N<4$
\cite{lowersusy,fronsd,kalkum,zaffa}.
\par
Here we follow the opposite viewpoint, namely starting
directly from $M$ theory we consider the
soliton solutions  of D=11 Supergravity that interpolate
between the other Freund Rubin vacua near a horizon and some flat
manifold
near spatial infinity. {\sl For each Freund Rubin manifold there
exists a
corresponding soliton. We name it a $G/H$ M--brane.} As we
show in the next section the crucial thing is the existence of
an Einstein $G$--invariant metric on $G/H$. These metrics were
explicitly constructed for all seven and four--dimensional coset
manifolds and can be utilized in the explicit derivation of the new
$11$--dimensional interpolating soliton metrics.
The next question is whether such classical solitons  are BPS
states, namely whether they admit suitable Killing spinors. The
answer is once again provided by the old results obtained
in the context of Kaluza Klein supergravity. What matters is
the number $N_{G/H}$ of (commuting) Killing spinors on
$G/H$ defined as the
solutions
of the following equation:
\begin{equation}
\left [ \Dcal^{G/H}_m + e \Gamma _m \right ] \eta =0
\label{kilghspi}
\end{equation}
where $\Dcal^{G/H}_m$ is the spinorial covariant derivative on $G/H$
calculated with respect to the $G$--invariant spin--connection and
$ \Gamma _m$ denotes the Dirac matrices in dimension
$ \mbox{dim}\frac{G}{H}$. The parameter $e$ is the Freund--Rubin
parameter, namely the vev of the $4$--form field strength which sets
also
the scale of anti de Sitter space. For all Freund Rubin manifolds
eq.\eqn{kilghspi} was thorougly analised in the eighties, the number
$N_{G/H}$ was determined and the solutions $\eta$ were explicitly
constructed. In the next section we show that each of such solutions
can be used to construct a Killing spinor for the corresponding
interpolating soliton. The new  Killing spinor is restricted, just
as
in the
case of ordinary Mp--branes, by the action of a projection operator
that
halves its $32$--components. Hence the conclusion is that for $G/H$
M--branes the number of preserved supersymmetries is given by the
following calculation:
\begin{eqnarray}
\# \mbox{supersymmetries in $G/H$ 2--brane} & = &
\frac{1}{2}\,  \times \, \frac{32}{8} \, \times
 N_{G/H}  = 2  \,
 N_{G/H} \label{nsusy7}\\
\# \mbox{supersymmetries in $G/H$ 5--brane} & = &
\frac{1}{2}\,  \times \, \frac{32}{4} \,
 \times N_{G/H}   \, = 4 \,   N_{G/H}
  \label{nsusy4}
\end{eqnarray}
In the above equation the factor $\frac{1}{2}$ accounts for the
aforementioned projection, the factors $\frac{32}{8}$ or
$\frac{32}{4}$ account  for the fact that spinor charges are counted
in units of either $8$--component spinors for $(G/H)_7$ or
$4$--component spinors for $(G/H)_4$. In any case the total number
of
spinor charges preserved by the $G/H$ M--brane is $1/2$ of the
number
of spinor charges preserved by the corresponding Freund--Rubin
solution. Indeed
the
Killing spinors of the Freund-Rubin vacua
are $N_{G/H}$ tensor products of a $4$--component spinor with an
$8$-component spinor for the $2$--brane case and of an
$8$--component
spinor with a $4$--component spinor in the $5$--brane case. This is
the familiar near horizon doubling of supersymmetries.
\vskip 0.1cm
{\sl Summarizing: all $G/H$ M--branes with
$ N_{G/H} \, > \, 0 $ are BPS states.}
\vskip 0.1cm
The bosonic isometry group of these classical solutions is given by
\begin{equation}
{\cal I}_{G/H-p-brane} \, = \, ISO(1,p) \, \otimes \, G
 \label{isometr2}
\end{equation}
which replaces eq.\eqn{isometr1}. Furthermore, in the case
$\left( \frac{G}{H} \right)_7$, recalling results that were
obtained in the early eighties \cite{pietricmspec}, we know that, if
the Freund Rubin coset manifold  admits $N_{G/H}$ Killing
spinors, then the structure of the isometry group $G$ is necessarily
the following one:
\begin{equation}
G \, = \, G^\prime \, \otimes \, SO\left( N_{G/H} \right)
\end{equation}
so that the factor $SO\left( N_{G/H} \right)$
can be combined with the isometry group $SO(2,3)$ of anti de Sitter
space
to produce the orthosymplectic algebra $Osp \left( N_{G/H} \vert 4
\right
)$. The same argument
leads to the conclusion that in the case $\left( \frac{G}{H}
\right)_4$,
the existence of $N_{G/H}$ Killing spinors should imply:
\begin{equation}
G \, = \, G^\prime \, \otimes \, Usp\left( N_{G/H} \right)
\end{equation}
In this way the factor $Usp\left( N_{G/H} \right)$ can be combined
with the anti de Sitter group $SO(2,6)$ into the ortosymplectic
algebra $Osp\left( 2,6\vert N_{G/H}\right)$.
\par
Therefore as the microscopic effective
action of ordinary Mp--branes  is invariant under transformations
of the
superconformal algebras \eqn{scgroup}, in the same way we can
conjecture that for $G/H$ Mp--branes the world--volume action should
have the following superconformal symmetries:
\begin{equation}
\begin{array}{ccc}
{\cal SC}_2^{G/H} & = & Osp \left( N_{G/H} \vert 4 \right) \, \times
\,
G^\prime \\
 \null & \null & \null \\
{\cal SC}_5^{G/H} & = & Osp \left( 2,6 \vert N_{G/H} \right) \,
\times
\,
G^\prime
\\
\end{array}
\label{scGH}
\end{equation}
\vskip 0.4cm
\noindent
 {\bf $\bullet$ Dimensional transmigration, anti de Sitter space
and solvable algebras }
\vskip 0.1cm
The key point in the above outlined developments is a mechanism that
we might describe as a dimensional transmigration. In Freund--Rubin
solutions the $11$--dimensional space is split in either $4+7$ or
$7+4$, the second number denoting the dimensions of the compactified
manifold and the first those of the effective space--time.
On the other hand, from the viewpoint of the Mp--brane solution the
dimensional split is either $11=3+8$ or $11=6+5$ the first number
denoting the world--volume dimensions, the second number the
dimensions of the transverse space. Hence the existing
relation between the superisometry group of the near horizon
geometry
and the superconformal symmetry of the
world--volume action involves the following dimensional
transmigration: anti de Sitter space $AdS_{p+2}$ looses one
of its $p+2$ dimensions and becomes the $d=p+1$ dimensional
world--volume of the $p$--brane. The lost dimension is
swallowed by the compact manifold $G/H$ that, by absorbing it,
becomes the transverse manifold to the $p$-brane.
The interpolation between two vacua performed by the M--brane
soliton
is nothing else but such a dimensional transmigration which occurs
smoothly while going from the horizon to spatial infinity.
This is a nice counting but it poses an obvious question.
How can we intrinsically characterize the $1$--dimensional
submanifold of
anti de Sitter space that can transmigrate to the transverse
manifold?
As we show in section \ref{solvbertot} and more extensively in
appendix \ref{mario}, the use of solvable Lie algebras answers this
question. Anti de Sitter space is a non compact  pseudo--riemanian
coset manifold:
\begin{equation}
 AdS_{p+2}  \, \equiv \, \frac{SO(2,p+1)}{SO(1,p+1)}
 \label{definisco}
\end{equation}
yet, in the same way as all the riemanian non--compact coset
manifolds, it can be identified with a solvable group manifold, that
is:
\begin{equation}
 AdS_{p+2}  \, =\, \exp \left [ Solv \right]
 \label{solvay}
\end{equation}
where $Solv$ denotes an appropriate $p+2$--dimensional solvable Lie
algebra.
The structure of this solvable algebra is simple: it contains a
$p+1$--dimensional {\it abelian ideal} ${\cal A}$ and a single
Cartan
generator ${\cal C}$. The solvable group parameters associated with
the abelian ideal ${\cal A}$ span the submanifold of $AdS_{p+2}$
that
can be viewed as the $p$--brane world volume. On the other hand the
$1$--dimensional submanifold generated by the Cartan operator ${\cal
C}$
is the one that performs the transmigration. Indeed naming $\rho$
this coordinate we find that it is nothing else but the the square
of
the radial coordinate \eqn{rdef}:
\begin{equation}
\rho = r^2
\label{agnisco}
\end{equation}
As we show in the appendix and in section \ref{solvbertot} the
solvable parametrization of anti de Sitter space is that which is
naturally provided by the Bertotti Robinson metric
\cite{bertotrobin}.
Originally,
Bertotti and Robinson introduced a metric in $4$--dimensions which
describes the tensor product of an anti de Sitter space $AdS_2$ with
a $2$--sphere $S^2$. Their metric can be easily generalized to
dimensions $D$ and describes the tensor product of an anti de
Sitter space $AdS_{p+2}$ with a sphere $S_{D-p-2}$. It reads as
follows:
\begin{equation}
ds^2\, = \,\rho^2 (-dt^2+d\vec{z}\cdot d\vec{z})+\rho^{-2}d\rho^2 \,
+ \,
d\Omega_{D-p-2}^2
\label{bertarob}
\end{equation}
where the last term $d\Omega_{D-p-2}^2$ is the invariant metric on
the
$S_{D-p-2}$ sphere, while the previous ones correspond to a
particular
parametrization of the anti de Sitter metric on $AdS_{p+2}$. This
parametrization is precisely the solvable one, the $p+1$ coordinates
$t,\vec{z}$ being those associated with the  abelian
ideal, while $\rho$ is associated with the Cartan generator.
\par
Upon the identification \eqn{agnisco},
the $D=11$ Bertotti Robinson metric \eqn{bertarob}  is the horizon
limit
($ r \, \to \, 0$) of the ordinary Mp-brane metric \eqn{protyp}.
The reason why the $S_{D-p-2}$ sphere emerges is the standard fact
that,
using polar coordinates, flat space $\IR^{D-p-1}$ can be viewed as a
sphere
fibration over the positive real line $\IR_+$. Indeed we can write
the familiar identity:
\begin{equation}
\, dX^I \,
dX^J \, \delta_{IJ} \, = \, dr^2 \, + \, r^2 \, d\Omega_{D-p-2}^2
\label{polar}
\end{equation}
The crucial observation for the derivation of $G/H$ M--branes is
that
as $D-p-1$-dimensional manifold transverse to the $p$--brane, rather
than a sphere fibration we can consider a $G/H$--fibration on
$\IR_+$, $G/H$ being a $D-p-2$--dimensional coset manifold. This
is made possible by the simultaneous fibered structure of anti de
Sitter space. In other words the base--manifold $\IR_+$ is shared in
the bulk of the soliton solution by both fibres, the world--volume
fibre and the transverse $G/H$ fibre. When we approach one of the
two
limits  $r \, \to \, 0$ or  $r \, \to \, \infty$ we reconstruct
either the anti de Sitter fibration or the transverse space
fibration.
\par
The solvable parametrization of anti de Sitter space is also the key
to understand the reinterpretation of the isometry superalgebras
\eqn{scGH} as superconformal algebras on the brane world--volume.
Although our analysis can be extended to the entire
supertransformations let us for the moment focus on the bosonic
ones.
 From the world--volume viewpoint the Cartan coordinate $\rho$
becomes
a scalar field that enters the generalized Born--Infeld action. Then
the {\it broken conformal transformations} found by Maldacena in
\cite{maldapasto} are nothing else but the ordinary action of the
isometry group $SO(2,p+1)$ on the $SO(2,p+1)/SO(1,p+1)$ coset
representative when the solvable parametrization is adopted.
This we explicitly verify in section \ref{conftrasformaz}.
\par
Our paper is organized as follows. In section \ref{leonard} we
derive the $G/H$ p--brane soliton solutions of D=11 supergravity.
In section \ref{BPSsusy} we discuss their Killing spinors and we
show that they are BPS states. In section \ref{solvbertot} we
analyse the solvable parametrization of anti de Sitter space
$AdS_4$ and we show that it leads to the Bertotti Robinson form of
the metric. In section \ref{conftrasformaz} we calculate the
Killing vectors representing the $SO(2,3)$ Lie algebra on the
solvable
coordinates and from them we exactly retrieve the form of Maldacena
broken conformal transformations. Section \ref{conclu} contains
our conclusions. Finally in appendix \ref{mario} we discuss the
generalization of our results to a generic $AdS_{p+2}$ space and
we show how the appropriate solvable Lie algebra can be constructed
using Iwasawa decomposition.
\section{Derivation of the $G/H$ M--brane solutions}
\label{leonard}
In this section we derive the $G/H$ M--brane solutions advocated in
the introduction, whose existence was originally proved in
\cite{duffetal}.
Such configurations are classical solutions of $D=11$ supergravity;
yet we find it convenient to start by revisiting the derivation
of $p$--brane solutions in a generic space--time dimension $D$.
So doing we can better illustrate the nature of the generalization
we propose. Indeed  the
assumption that transverse space has the topology of $\IR^{D-p-1}$
nowhere appears to be necessary.
\par
In the most general setting $p$--branes are thought of as solutions
of
$N$--extended supergravity theories. Hence the relevant bosonic
lagrangian involves a number $f$ of $n$--form field strengths
$F^\Lambda_{M_1\dots M_n}=n\part_{[M_1} A^\Lambda_{M_2\dots M_n]}$
($M_i=0,\ldots,D-1$, $\Lambda=1,\dots, f$) and a number $s$ of
scalar
 fields $\phi^i$
($i=1,\dots,s$), the relation between the number of space--like
dimensions of the brane and the degree of the field--strength being
\begin{equation}
n=p+2
\label{nisp+2}
\end{equation}
In this setting the elaborated geometry of the scalar sector and its
relation with the group of duality transformations plays an
important
role. One example of this is provided by the discussion of the most
general $0$--brane solutions (black--holes) in $N=8$ supergravity
where $f = 28$ and $s=70$ \cite{noi3}.
\subsection{The general $p$--brane action in $D$--dimensions}
It is important to take into account all
the field strengths and all the scalar fields in order
to study the orbits of the U--duality group and the moduli
dependence
of the solution. However if we are interested in the space--time
structure of the $p$--brane soliton it is sufficient to restrict
our attention to a lagrangian of the following type
\cite{mbrastelle}
:
\begin{equation}
I=\int d^D x \sqrt{-g}~ [R-{1\over 2} \nabla_M \phi \nabla^M \phi-
{1
\over 2
n!} e^{a \phi} F^2_{[n]}]
\label{gendact}
\end{equation}
involving only the metric $g_{MN}$, a single scalar $\phi$ (the
dilaton)
and a single $(n-1)$-form gauge potential $A_{[n-1]}$ with field
strength
$F_{[n]}$ ( the parameter $a$ is the scalar coupling).
{}From the point of view of the complete supergravity theory, eq.
\eqn{gendact}
corresponds to U--rotate the field strength vector to a standard one
with a single non--vanishing component and truncate the action to
such a sector. Similarly  $\phi$ denotes the combination
of scalars that couples to the selected field strength.
\par
The field equations derived from \eqn{gendact} have the following
form:
\begin{eqnarray}
& &R_{MN}={1\over 2} \partial_M \phi \partial_N \phi
+S_{MN}\label{feq1}\\
& &\nabla_{M_1} (e^{a\phi} F^{M_1 ...M_n})=0 \label{feq2}\\
& &\square \phi = {a\over 2n!} F^2 \label{feq3}
\end{eqnarray}
where $S_{MN}$ is the energy-momentum tensor of  the $n$-form $F$:
\begin{equation}
S_{MN}={1 \over 2(n-1)!} e^{a\phi}  [F_{M...} F_{N}^{~...} - {n-1
\over n(D-2)}
F^2 g_{MN}]
\end{equation}
\subsection{The $G/H$ electric $p$-brane ansatz}
Motivated by the arguments discussed in section \ref{intro}
we search for solutions of eq.s\eqn{feq1},\eqn{feq2},\eqn{feq3}
of the form:
\begin{eqnarray}
& &ds^2=e^{2A(r)} dx^{\mu} dx^{\nu} \eta_{\mu\nu} + e^{2B(r)}[dr^2
+r^2
\la^{-2} ds^2_{G/H}] \label{Ansatz1}\\
& &A_{\mu_1 ...\mu_d}=\epsilon_{\mu_1...\mu_d} e^{C(r)}
\label{Ansatz2}\\
& &\phi=\phi(r) \label{Ansatz3}
\end{eqnarray}
where
\begin{enumerate}
\item  $ \lambda $ is a constant parameter with
the dimensions of  length.
\item The $D$ coordinates $X^M$ are split as follows:
$X^M =(x^{\mu},r,y^m)$, $\eta^{MN}=\mbox{diag}(-,+++...)$
\item $\mu= 0,\dots ,d-1 $ runs on the  p-brane world-volume
($d=p+1$)
\item
$\bullet$ labels  the $r$ coordinate
\item
$m=d+1,\dots, D-1$ runs on some $D-d-1$-dimensional compact
coset manifold $G/H$, $G$ being a compact Lie group and $H\subset G$
a closed Lie subgroup.
\item $ds^2_{G/H}$ denotes a $G$--invariant metric on the above
mentioned coset manifold.
\end{enumerate}
Eq.s \eqn{Ansatz1},\eqn{Ansatz2},\eqn{Ansatz3} provide the $G/H$
generalization of the standard electric $p$--brane ansatz
extensively
considered in the literature (see for instance \cite{mbrastelle}).
Indeed, the only difference with the ordinary case is that we have
replaced the invariant metric $ds^2_{S^{D-d-1}}$ on a sphere
$S^{D-d-1}$ by the more general coset manifold metric  $ds^2_{G/H}$.
Applied to the case of $D=11$ supergravity, the electric ansatz
will produce both $G/H$ M2--brane and M5--brane solutions.
\par
On the other hand it is well known that
for ordinary branes there exists also a magnetic solitonic
ansatz. The $G/H$ generalization of such a magnetic ansatz is
straightforward but we do not dwell on it in this paper,
leaving a more in depth analysis for a future publication.
\par
As anticipated in the introduction the isometry group of the
field configuration introduced by the electric ansatz
\eqn{Ansatz1},\eqn{Ansatz2},\eqn{Ansatz3} is given by the group
$ {\cal I}_{G/H-p-brane}$ defined in eq.\eqn{isometr2}.
\subsubsection{The Vielbein}
In order to prove that the ansatz \eqn{Ansatz1}-\eqn{Ansatz3}
is a solution of the field equations
it is necessary to calculate the corresponding vielbein,
spin--connection and curvature tensors. We use the
convention that tangent space indices are underlined. Then the
vielbein components relative to the
ansatz (\ref{Ansatz1}) are:
\begin{eqnarray}
& &\Ef{\mu}=e^Adx^{\mu};~\Ef{\bu} =e^Bdr;~~\Ef{m}=e^B r \la^{-1}
\Efb{m}; \\
& &g_{\mu\nu}=e^{2A}
\eta_{\mu\nu};~~g_{\bu\bu}=e^{2B},~~g_{mn}=e^{2B} r^2
\la^{-2}\gb_{mn}
\end{eqnarray}
with $\Eb^{\mun} \equiv G/H$ vielbein and  $\gb_{mn}$ $\equiv $
$G/H$
metric.
\subsubsection{The spin connection}
The Levi--Civita spin--connection on our $D$--dimensional manifold
is defined as the solution of the vanishing torsion equation:
\begin{equation}
d\Ef{M} + \omef{M}{N} \we \Ef{N}=0
\label{zerotors}
\end{equation}
Solving eq.\eqn{zerotors} explicitly we obtain the spin--connection
components:
\begin{equation}
\omefu{\mu}{\nu}=0,~~\omefu{\mu}{\bu}=e^{-B} A'
\Ef{\mu},~~\omefu{\mu}{n}=0,~~
\omefu{m}{n}=\omefub{m}{n},~~\omefu{m}{\bu}=e^{-B}
(B'+r^{-1})\Ef{m}.
\end{equation}
where $A' \equiv \part_\bu A$ etc.  and $\omefub{m}{n}$ is the spin
connection of the $G/H$ manifold.
\subsubsection{The Ricci tensor}
 From the definition of the curvature 2-form :
\begin{equation}
\Rfu{M}{N}=d\omefu{M}{N}+\omef{M}{S} \we \omefu{S}{N}
\end{equation}
we find the Ricci tensor components:
\begin{eqnarray}
& &R_{\mu\nu}=-{1\over 2} \eta_{\mu\nu} e^{2(A-B)}  [A''+d
(A')^2+\dt
A'B' +
(\dt+1) r^{-1} A']\\
& &R_{\bu\bu}=-{1\over 2}
[d(A''+(A')^2-A'B')+(\dt+1)(B''+r^{-1}B')]\\
& &R_{mn}=-{1\over 2} \gb_{mn}  {r^2\over \la^2}
[dA'(B'+r^{-1})+r^{-1}B'+B''+\dt (B'+r^{-1})^2]+ \Rb_{mn}
\end{eqnarray}
where $\Rb_{mn}$ is the Ricci tensor of $G/H$ manifold, and $\dt
\equiv
D-d-2$.
\subsubsection{The field equations}
Inserting the electric ansatz into the field eq.s \eqn{feq1} yields:
\begin{eqnarray}
& & A''+d(A')^2+\dt A'B' + (\dt+1) A' r^{-1} = {\dt \over 2(D-2)}
S^2 \label{einstein1}\\
& & d[A''+(A')^2-A'B']+(\dt+1)[B'' + r^{-1}B']={\dt \over 2(D-2)}
S^2-{
(\phi')^2\over 2} \label{einstein2}\\
& & \gb_{mn} [dA'(B'+r^{-1})+r^{-1}B'+B''+\dt (B'+r^{-1})^2]-2
\Rb_{mn}=\nonumber\\
& & ~~~~~~~~~~~~~~~~~~~~~~~~~~~~~~~~~~~~~~~
{}~~~~~~~~~~~~~~~~~~~-{d\over 2(D-2)} \gb_{mn} S^2 \label{einstein3}
\end{eqnarray}
while eq.s (\ref{feq2})-(\ref{feq3}) become:
\begin{eqnarray}
C''+ (\dt+1) r^{-1} C' + (\dt B'-d A'+C'+a \phi') C'=0
\label{maxwell}\\
\phi''+ (\dt+1) r^{-1} \phi' + [dA'+\dt B'] \phi'= -{a\over 2} S^2
\label{scalar}
\end{eqnarray}
with
\begin{equation}
S \equiv e^{{1\over 2} a \phi +C -dA} C'
\end{equation}
\subsection{Construction of the BPS Killing spinors in the case
of $D=11$ supergravity}
\label{BPSsusy}
At this point we specialize our analysis to the case
of $D=11$ supergravity, whose action in the bosonic sector reads:
\begin{equation}
I_{11}=\int d^{11}x \sqrt{-g} ~(R-{1\over 48} F^2_{[4]} ) + {1\over
6} \int
F_{[4]} \we
F_{[4]} \we A_{[3]}
\label{elevenaction}
\end{equation}
and we look for the further restrictions imposed on the electric
ansatz by the requirement that the solutions should preserve a
certain amount of supersymmetry. This is essential for our goal
since we are interested in $G/H$ M--branes that are BPS saturated
states and the  BPS condition requires the existence of Killing
spinors.
\par
As discussed  in ref. \cite{mbrastelle},  the above action does not
fall exactly in the general class of actions of type  \eqn{gendact}.
Nevertheless, the results of sections 2.1 and 2.2 still apply:
indeed it is straightforward to verify that the FFA term in  the
action
\eqn{elevenaction} gives no contribution to the field equations
once the electric or magnetic ansatz are implemented.
Moreover no scalar fields are present in \eqn{elevenaction}:
this we handle by simply  setting  to zero the scalar coupling
parameter $a$.
\par
Imposing that the ansatz solution admits Killing spinors
allows to simplify the field equations drastically.
 \par
We recall the supersymmetry
transformation for the gravitino:
\begin{equation}
\de \psi_M = \Dt_M \epsilon
\label{gravitin1}
\end{equation}
with
\begin{equation}
\Dt_M = \part_M + {1\over 4} \om_M^{~~AB} \Ga_{AB} -
{1 \over 288} [\Ga^{PQRS}_{~~~~M} + 8 \Ga^{PQR} \de^S_M] F_{PQRS}
\label{gravitin2}
\end{equation}
Requiring that setting $\psi_M=0$  be consistent with the
existence
of residual supersymmetry yields:
\begin{equation}
{\de \psi_M }_{|\psi=0}=\Dt_M \epsilon=0
\label{Kspinor}
\end{equation}
Solutions $\epsilon (x,r,y)$ of the above equation are {\sl Killing
spinor} fields on the bosonic background described by our ansatz.
\par
In order to discuss the solutions of \eqn{Kspinor}
we adopt the following tensor product realization
of the ($32 \times 32$) $SO(1,10)$ gamma matrices:
\begin{equation}
\Ga_A=[\ga_{\mu} \otimes \bfone_8, \ga_3 \otimes \bfone_8, \ga_5
\otimes
\Ga_m]
\label{gambas}
\end{equation}
The above basis \eqn{gambas} is well adapted to our (3+1+7) ansatz.
The $\ga_{\mu} ~(\mu=0,1,2,3)$ are usual $SO(1,3)$ gamma
matrices,  $\ga_5 = i\ga_0 \ga_1 \ga_2 \ga_3$, while $\Ga_m$ are
$8\times 8$
gamma matrices realizing the  Clifford algebra
of $SO(7)$.  Thus for example $\Ga_{\bu}=\ga_3 \otimes \bfone_8$.
 \par
Correspondingly, we split the D=11  spinor $\epsilon$ as follows
\begin{equation}
\epsilon = \varepsilon  \, \otimes \, \eta (r,y)
\end{equation}
where $ \varepsilon $ is an $SO(1,3)$ constant spinor, while the
$SO(7)$ spinor $\eta $, besides the dependence on the internal $G/H$
coordinates
$y^m$, is assumed to depend also on the radial coordinate $r$.
Note the difference with respect to Kaluza Klein supersymmetric
compactifications where $\eta$ depends only  on $y^m$.
Computing $\Dt$ in the ansatz background yields:
\begin{eqnarray}
& & \Dt_{\mu}=\part_{\mu} + {1\over 2} e^{-B-2A}\ga_{\mu} \ga_3
[e^{3A} A' -
 {i\over 3} e^C C' \ga_3 \ga_5] \otimes \bfone_8 \nonumber \\
& &\Dt_{\bu} = \part_r + {i\over 6} e^{-3A} C' e^C \ga_3 \ga_5
\otimes \bfone_8 \nonumber \\
& &\Dt_m=\Dcal^{G/H}_m+{r\over 2 \la} [(B' + r^{-1}) i\ga_3\ga_5+
{1\over 6} e^{C-3A} C']\otimes \Ga_m
\label{splinter}
\end{eqnarray}
where all $\ga_{\mu}$, $\Ga_m$ have tangent space indices.
The Killing spinor equation  $\Dt_{\mu} \epsilon=0$  becomes
equivalent to:
\begin{equation}
(1_4-i \ga_3\ga_5)\epsi = 0 ; ~~~3e^{3A} A'=e^CC'
\label{condominio}
\end{equation}
Thus half of the components of the 4-dim spinor $\epsi$ are
projected
out.
Moreover the second equation is solved by $C=3A$.
Considering next $\Dt_{\bu} \epsilon=0$ leads to the equation (where
we have used  $C=3A$):
\begin{equation}
\part_r \eta+{1\over 6} C' \eta=0
\end{equation}
whose solution is
\begin{equation}
\eta(r,y)=e^{-C(r)/6} \eta_{\ci}(y)
\end{equation}
Finally, $\Dt_m \epsilon=0$ implies
\begin{eqnarray}
& & B=-{1\over 6} C+const.\\
& & [\Dcal^{G/H}_m + {1\over 2\la}  \Ga_m] \eta_{\ci}=0
\label{meraviglia}
\end{eqnarray}
Eq.\eqn{meraviglia} deserves attentive consideration. If we identify
the Freund--Rubin parameter as:
\begin{equation}
 e \equiv {1\over 2\la}
\end{equation}
then eq.\eqn{meraviglia} is nothing else but the Killing spinor
equation for a $G/H$ spinor that one encounters while discussing
the residual supersymmetry of Freund--Rubin vacua. The solutions of
this
equation
have been exhaustively studied in  the old literature on
Kaluza--Klein supergravity (see \cite{castdauriafre} for a
comprehensive review) and are all known.
\par
In this way we have explicitly verified what was mentioned in the
introduction, namely that the number of $BPS$ Killing spinors
admitted by the $G/H$ M--brane solution is $N_{G/H}$, i.e.
the number of Killing spinors admitted by the corresponding
Freund--Rubin vacuum.
\subsection{M-brane solution}
To be precise the Killing spinors of the previous section are
admitted by a
configuration
that has still to be shown to be a complete solution of the field
equations. To prove this is immediate.
Setting $D=11,d=3,\dt=6$, the scalar coupling parameter $a=0$, and
using the
relations
$C=3A,B=-C/6+const.=-A/2+const.$ we have just deduced,  the field
equations
\eqn{einstein1}, \eqn{einstein2}, \eqn{einstein3}  become:
\begin{eqnarray}
& &A''+ 7 r^{-1} A'={1\over 3} S^2 \\
& & (A')^2={1\over 6} S^2\\
& &\Rb_{mn} = {3\over  \la^2} \gb_{mn} \label{Einstein}
\end{eqnarray}
Combining the first two equations to eliminate $S^2$ yields:
\begin{equation}
\nabla^2 A - 3 (A')^2 \equiv A''+{7 \over r} A' - 3 (A')^2=0
\label{afieldeq}
\end{equation}
or:
\begin{equation}
\nabla^2 e^{-3A} = 0 \label{nablaA}
\end{equation}
whose solution is:
\begin{equation}
e^{-3A(r)}= H(r)=1+{k\over r^6}
\end{equation}
We have chosen the integration constant such that $A(\infty)=0$.
The functions $B(r)$ and $C(r)$ are then given by  $B=-A/2$ (so that
$B(\infty)=0$) and $C=3A$.
Finally, after use of $C=3A$, the F-field equation
\eqn{maxwell} becomes equivalent to  \eqn{afieldeq}.
The equation \eqn{nablaA} determining the radial dependence of the
function
$A(r)$
(and consequently of $B(r)$ and $C(r)$)
 is the same here as in the case
of ordinary branes, while to solve eq.\eqn{Einstein} it suffices to
choose for the manifold $G/H$ the $G$--invariant Einstein metric.
Each of the Freund--Rubin cosets admits such an Einstein metric
which
was also constructed in the old Kaluza--Klein supergravity
literature
(see \cite{KKwarncastel,castdauriafre})
\par
Summarizing:  for  $D=11$ supergravity  the field equations are
solved by the ansatz \eqn{Ansatz1}, \eqn{Ansatz2}
where the $A,B,C$ functions are
\begin{eqnarray}
  A(r) & = & -\frac{\dt}{18} \, \ln
  \left(1+\frac{k}{r^{\dt}}\right) = -{1\over 3} \ln
\left(1+\frac{k}{r^{6}}\right)\nonumber \\
  B(r) & = & \frac{d}{18} \, \ln
  \left(1+\frac{k}{r^{\dt}}\right) = {1\over 6}  \ln
\left(1+\frac{k}{r^{6}}\right)\nonumber \\
  C(r) &=& 3 \, A(r)
\end{eqnarray}
 displaying the same $r$--dependence as the ordinary M--brane
solution
\eqn{protyp}.
\par
In this way we have illustrated the existence of $G/H$
M--brane solutions (cf. \cite{duffetal}). Table 1 displays the
Freund--Rubin cosets with non vanishing $N_{G/H}$. Each of them
is associated to a BPS saturated M--brane. The notations are as in
ref.s \cite{KKwarncastel}, \cite{castdauriafre}.
\par
\vskip 0.3cm
\begin{table}[ht]
\begin{center}
\caption{Supersymmetric Freund Rubin Cosets with Killing spinors}
\label{tavola}
\vskip .3cm
 \begin{tabular}{|c|c|c|c|}\hline
  G/H &   $G$     &   $H$   &   $N_{G/H}$    \\
 \hline
 \hline
{}&{}&{}&{}\\
 $S^7$ &  $SO(8)$ &   SO(7) &   8             \\
 \hline
 squashed $S^7$ & $SO(5) \times SO(3)$ & $ SO(3)\times SO(3)$& 1\\
 \hline
 $M^{ppr}$ & $SU(3) \times SU(2) \times U(1)$ & $SU(2) \times
U(1)^2$
& 2\\
\hline
$N^{010}$ & $SU(3) \times SU(2)$ & $ SU(2)\times U(1)$& 3\\
\hline
$N^{pqr}$ & $SU(3) \times U(1)$ & $U(1)^2$& 1\\
\hline
$Q^{ppp}$ & $SU(2)^3 $ & $U(1)^3$& 2\\
\hline
$B_{irred}^7$ & $SO(5)$ & $ SO(3)_{max}$& 1\\
\hline
$V_{5,2}$ & $SO(5)) \times U(1)$ & $ SO(3)\times U(1)$& 2\\
\hline
 \end{tabular}
 \end{center}
\end{table}
\section{ $AdS_4$ parametrization and the Bertotti Robinson
metric}
\label{solvbertot}
In this section we consider the explicit example of the
4--dimensional anti de Sitter space:
\begin{equation}
AdS_4 \, \equiv \, \frac{SO(2,3)}{SO(1,3)}\, .
\label{ads4}
\end{equation}
Relying on the algebraic derivation explained in appendix
\ref{mario}
we
claim that this coset manifold can be identified with the
exponential
of
a $4$--dimensional solvable Lie algebra $Solv_4$. The complex form
of
the
$SO(2,3)$ Lie algebra is $B_2$ and the root system is composed by
the
eight roots:
\begin{equation}
\pm \epsilon_1 \pm\epsilon_2 \quad ; \quad \pm \epsilon_1 \quad ,
\quad \pm \epsilon_2
\label{rutte}
\end{equation}
where $\epsilon_i$ denote the unit vectors in a Euclidean
two--dimensional space. Adopting the standard notation   $E_\alpha $
for the step operator  associated with the root $\alpha$ and
${\cal H}_\alpha$
for the Cartan generator obtained by commuting $E_\alpha $ with
$E_{-\alpha }$, the results of the appendix yield the
following conclusion. The solvable Lie algebra $Solv \left ( AdS_4
\right)$
generating $4$--dimensional anti de Sitter
space  is spanned by the following three  nilpotent operators
\begin{equation}
{\cal T}_ \perp \, \equiv \, E_{\epsilon_2} \quad ; \quad {\cal T}_-
 \, \equiv \,
E_{\epsilon_2 + \epsilon_1} \quad ;
\quad {\cal T}_+ \, \equiv \,  E_{\epsilon_2 - \epsilon_1}
\label{abetrasl}
\end{equation}
plus the following non--compact Cartan generator
\begin{equation}
{\cal C} \, \equiv \, H_{ \epsilon_2}
\end{equation}
The matrix realization of these generators in the ${\underline {\bf
5}}$
of $SO(2,3)$ is:
\begin{eqnarray}
{\cal C}
\,&=&\,\left( \matrix{ 0 & 0 & 0 & 0 & 0 \cr 0 & 0 & 0 & 1 & 0
   \cr 0 & 0 & 0 & 0 & 0 \cr 0 & 1 & 0 & 0 & 0
   \cr 0 & 0 & 0 & 0 & 0 \cr  }
\right)\,\,
   {\cal T}_\perp\,=\,\left(
\matrix{ 0 & 0 & 0 & 0 & 0 \cr 0 & 0 & 0 & 0 &
  {\frac{1}{{\sqrt{2}}}} \cr 0 & 0 & 0 & 0 & 0 \cr 0
   & 0 & 0 & 0 & {\frac{1}{{\sqrt{2}}}} \cr 0 &
  {\frac{1}{{\sqrt{2}}}} & 0 & -{\frac{1}{{\sqrt{2}}}}
   & 0 \cr  }
\right)\nonumber\\
{\cal T}_-\,&=&\,\left(
\matrix{ 0 & -{\frac{1}{2}} & 0 & {\frac{1}{2}} & 0
   \cr {\frac{1}{2}} & 0 & -{\frac{1}{2}} & 0 & 0 \cr
  0 & -{\frac{1}{2}} & 0 & {\frac{1}{2}} & 0 \cr
  {\frac{1}{2}} & 0 & -{\frac{1}{2}} & 0 & 0 \cr 0 &
  0 & 0 & 0 & 0 \cr  }
\right)
\,\,{\cal T}_+\,=\,\left(
\matrix{ 0 & -{\frac{1}{2}} & 0 & {\frac{1}{2}} & 0
   \cr {\frac{1}{2}} & 0 & {\frac{1}{2}} & 0 & 0
   \cr 0 & {\frac{1}{2}} & 0 & -{\frac{1}{2}} & 0
   \cr {\frac{1}{2}} & 0 & {\frac{1}{2}} & 0 & 0
   \cr 0 & 0 & 0 & 0 & 0 \cr  }\
\right)
\label{tpm}
\end{eqnarray}
The reason for eq.\eqn{abetrasl} is that
 the parameter associated with ${\cal T}_\perp, {\cal T}_-, {\cal
T}_+$ will
be
respectively
interpreted
 as the transverse  and   light--cone coordinates
 on the 2--brane world volume. This will be manifest
 at the end of our calculations. For the time being
 just take these as convenient names given to the solvable
 Lie algebra generators. Using such a notation we write the
coset representative in the following way:
\begin{eqnarray}
L(a,x ,t,w) &=&  \tau (x,t,w)\, S(a)  \nonumber\\
S(a) & \equiv &  \exp\left[ -a \, {\cal C}\right] \nonumber\\
\tau(x,t,w) & =& \exp \left[ \sqrt{2} \, x \, {\cal T}_\perp +
\,
(t-w) \, {\cal T}_- \, +  \, (t+w)
\, {\cal T}_+ \right ]
\label{cosettus}
\end{eqnarray}
By explicit evaluation we find:
\begin{equation}
S(a) \, = \, \left(
\matrix{ 1 & 0 & 0 & 0 & 0 \cr 0 & \cosh[a] & 0 &
  -\sinh[a] & 0 \cr 0 & 0 & 1 & 0 & 0 \cr 0 &
  -\sinh[a] & 0 & \cosh[a] & 0 \cr 0 & 0
   & 0 & 0 & 1 \cr  }
\right)
\label{samat}
\end{equation}
and
\begin{equation}
 \tau(x,t,w) \, = \,
 \left(
\matrix{ 1 & -t & 0 & t & 0 \cr t & {{2 - {t^2} + {w^2} +
{x^2}}\over 2}
& w
   & {{{t^2} - {w^2} - {x^2}}\over 2} & x \cr 0 & w & 1
&
-w & 0
\cr t &
  {{-{t^2} + {w^2} + {x^2}}\over 2} & w & {{2 + {t^2} - {w^2}
-
{x^2}}\over 2}
   & x \cr 0 & x & 0 & -x & 1 \cr  }
\right)
\label{taumat}
\end{equation}
Then it is straightforward to calculate the
 left invariant $1$--form and one obtains:
\begin{equation}
\Omega= L^{-1} \, d L = \left(
\matrix{ 0 & -\left( {\it dt}\,{e^a} \right)  & 0 & {\it dt}\,{e^a}
&
0 \cr
  {\it dt}\,{e^a} & 0 & {\it dw}\,{e^a} & -{\it da} & {\it
dx}\,{e^a} \cr
0 &
  {\it dw}\,{e^a} & 0 & -\left( {\it dw}\,{e^a} \right)  & 0 \cr
  {\it dt}\,{e^a} & -{\it da} & {\it dw}\,{e^a} & 0 & {\it
dx}\,{e^a} \cr
0 &
  {\it dx}\,{e^a} & 0 & -\left( {\it dx}\,{e^a} \right)
& 0 \cr  }
\right)
\label{leftinv}
\end{equation}
With these results we are now in a position to calculate the
vielbein,
the spin connection and the curvature of our anti de Sitter space in
the solvable parametrization.
\par
To this effect it suffices to write a standard basis of generators
for the $SO(2,3)$ Lie algebra singling out the $\IK$ coset
orthogonal
subspace from the $\IH \equiv SO(1,3)$ subalgebra.
\par
First we recall that in our convention the $SO(2,3)$ group is given
by the set of $5 \times 5$ matrices that leave invariant the
following diagonal metric:
\begin{equation}
\eta= \mbox{diag} \, \left( -,-,+,+,+ \right) \label{etametric}
\end{equation}
Written in standard form the $SO(2,3)$ Lie algebra is as follows:
\begin{eqnarray}
\left[ M^{AB} \, ,\,  M^{CD} \right] & = & - \eta^{AC} \, M^{BD}  \,
+ \,  \eta^{BC} \, M^{AD}  \,+ \,  \eta^{AD} \, M^{BC}
\,- \,  \eta^{BD} \, M^{AC} \nonumber\\
M^{AB}& =& - M^{BA} \, \quad ; \quad A,B= 1, \dots , 5
\end{eqnarray}
Furthermore the Lorentz subalgebra $SO(1,3)$ we have chosen is given
by the
subset of
$SO(2,3)$ Lie algebra matrices that are of the following form:
\begin{equation}
\left( \matrix{ \star & 0 & \star & \star &\star  \cr
                  0   & 0 & 0     &  0    & 0     \cr
                 \star & 0 & \star & \star &\star  \cr
                 \star & 0 & \star & \star &\star  \cr
                 \star & 0 & \star & \star &\star  \cr }
 \right)
 \label{lorentztype}
\end{equation}
Correspondingly we can write the orthogonal decomposition of the
$SO(2,3)$ Lie algebra:
\begin{equation}
SO(2,3)=\IH_{SO(1,3)}\, \oplus\, \IK
\label{ortdecompo}
\end{equation}
where the subalgebra $\IH_{SO(1,3)}$ is spanned by the three Lorentz
boosts $N_1, N_2, N_3$ and the three angular momenta $J_1,J_2, J_3$.
We list
below the explicit form of these generators and their correspondence
with the
generators $M^{AB}$:
\begin{equation}
 \begin{array}{cccc}
 M^{1,3} \equiv N_1=& \left(\matrix{ 0 & 0 & 1 & 0 & 0 \cr 0 & 0 & 0
& 0 & 0
\cr 1 & 0 & 0 & 0 & 0 \cr 0
   & 0 & 0 & 0 & 0 \cr 0 & 0 & 0 & 0 & 0 \cr  } \right)
   & M^{3,4} \equiv J_1= &\left(
\matrix{ 0 & 0 & 0 & 0 & 0 \cr 0 & 0 & 0 & 0 & 0 \cr 0 & 0 & 0 & 1 &
0 \cr 0
   & 0 & -1 & 0 & 0 \cr 0 & 0 & 0 & 0 & 0 \cr  }  \right) \\
M^{1,4} \equiv N_2=&\left(
\matrix{ 0 & 0 & 0 & 1 & 0 \cr 0 & 0 & 0 & 0 & 0 \cr 0 & 0 & 0 & 0 &
0 \cr 1
   & 0 & 0 & 0 & 0 \cr 0 & 0 & 0 & 0 & 0 \cr  } \right)
& M^{3,5} \equiv J_2= & \left(
\matrix{ 0 & 0 & 0 & 0 & 0 \cr 0 & 0 & 0 & 0 & 0 \cr 0 & 0 & 0 & 0 &
1 \cr 0
   & 0 & 0 & 0 & 0 \cr 0 & 0 & -1 & 0 & 0 \cr  } \right)
\\
M^{1,5} \equiv N_3=&
  \left(
\matrix{ 0 & 0 & 0 & 0 & 1 \cr 0 & 0 & 0 & 0 & 0 \cr 0 & 0 & 0 & 0 &
0 \cr 0
   & 0 & 0 & 0 & 0 \cr 1 & 0 & 0 & 0 & 0 \cr  } \right)
&M^{4,5} \equiv J_3= &
\left(
\matrix{ 0 & 0 & 0 & 0 & 0 \cr 0 & 0 & 0 & 0 & 0 \cr 0 & 0 & 0 & 0 &
0 \cr 0
   & 0 & 0 & 0 & 1 \cr 0 & 0 & 0 & -1 & 0 \cr  }  \right)
\\
 \end{array}
 \label{nnnjjj}
 \end{equation}
 On the other hand the orthogonal complement $\IK$ is spanned by the
 following four generators:
 \begin{equation}
\begin{array}{cccc}
M^{2,1} \equiv \Pi_0 = & \left(
\matrix{ 0 & 1 & 0 & 0 & 0 \cr -1 & 0 & 0 & 0 & 0 \cr 0 & 0 & 0 & 0
&
0 \cr 0
   & 0 & 0 & 0 & 0 \cr 0 & 0 & 0 & 0 & 0 \cr  }
   \right)
&M^{2,3} \equiv \Pi_1 = &\left(
\matrix{ 0 & 0 & 0 & 0 & 0 \cr 0 & 0 & 1 & 0 & 0 \cr 0 & 1 & 0 & 0 &
0 \cr 0
   & 0 & 0 & 0 & 0 \cr 0 & 0 & 0 & 0 & 0 \cr  }  \right)
\\
M^{2,4} \equiv \Pi_2 = & \left(
\matrix{ 0 & 0 & 0 & 0 & 0 \cr 0 & 0 & 0 & 1 & 0 \cr 0 & 0 & 0 & 0 &
0 \cr 0
   & 1 & 0 & 0 & 0 \cr 0 & 0 & 0 & 0 & 0 \cr  }
   \right)
&
M^{2,5} \equiv \Pi_3 = & \left(
\matrix{ 0 & 0 & 0 & 0 & 0 \cr 0 & 0 & 0 & 0 & 1 \cr 0 & 0 & 0 & 0 &
0 \cr 0
   & 0 & 0 & 0 & 0 \cr 0 & 1 & 0 & 0 & 0 \cr  }  \right)
 \\
\end{array}
\label{vielgene}
\end{equation}
Ordering the ten generators in a ten-vector as follows:
\begin{equation}
T_A \equiv \left \{ \Pi_0,\Pi_1,\Pi_2,\Pi_3,N_1,N_2,N_3,J_1,J_2,J_3
\right\}
\quad , \quad A=1,\dots , 10
\label{ordinamento}
\end{equation}
we find that they are trace-orthogonal according to:
\begin{equation}
 \mbox{Tr}\left( T_A, T_B \right)  =  {\bf k}_{AB}= \mbox{diag}
(-2,2,2,2,2,2,2,-2,-2,-2)
 \label{Killingmet}
\end{equation}
so that, by taking  traces, we can easily project the left invariant
$1$--form
\eqn{leftinv} along
the subspace $\IK$ or the subalgebra $\IH$ .
Such projections yield the vierbein and the spin connection of anti
de Sitter space, respectively. Let us begin with the calculation of
the
vielbein.
By definition we have:
\begin{eqnarray}
V^0 & = & -\frac{1}{2} \, \mbox{Tr} \left(  \, \Pi_0 \,  \Omega
\right)
\nonumber\\
V^i & = & \frac{1}{2} \, \mbox{Tr} \left(  \, \Pi_i \,  \Omega
\right)  \quad ,
\quad i=1,2,3
\end{eqnarray}
and we immediately obtain:
\begin{equation}
\left ( \matrix{ V^0 \cr  V^1 \cr  V^2 \cr  V^3 \cr }\right )\,=\,
  \left( \matrix{
  {\it dt}\,{e^a} \cr  {\it dw}\,{e^a} \cr
   -{\it da}\cr {\it dx}_\perp \,{e^a} \cr } \right)
\end{equation}
setting:
\begin{equation}
\rho \equiv e^a
\label{radius}
\end{equation}
and calculating the metric we obtain:
\begin{eqnarray}
ds^2 & \equiv & -V^0 \otimes V^0 \, + \, V^1 \otimes V^1 \, + \,
V^2 \otimes V^2 \, + \,
V^3 \otimes V^3 \nonumber\\
& = &  \rho^2 \, \left(-dt^2\,+\, dw^2\, + \, dx^2 \right) \,
+
\,
\rho^{-2} \, d\rho^2
\end{eqnarray}
which is the anti de Sitter metric in Bertotti Robinson form.
\section{The $SO(2,3)$ transformation rules}
\label{conftrasformaz}
Given the coset parametrization in terms of the solvable Lie algebra
parameters $( \rho, x  , t, w ) $
we can work out the explicit form of the Killing vectors
representing
the infinitesimal action of  $SO(2,3)$ on a general function of
$y^a\equiv \{ \rho,x,t,w\}$. We rely on the general formula
\cite{castdauriafre}:
\begin{equation}
T_A \, L(y) \, = \, k^a_A \, \frac{\partial}{\partial y^a} \, L(y)
\,
- \, L(y) \, T_i \, W_A^i(y)
\label{mastformul}
\end{equation}
where
\begin{equation}
\delta y^a \, \equiv \, \epsilon^A \, k^a_A (y)
\label{deltay}
\end{equation}
defines the Killing vectors, $\epsilon^A$ ($A=1,\dots , \mbox{dim}\,
G$)
are the Lie algebra parameters, $T_A$ and $T_i$ being the
generators of the full Lie algebra $G = SO(2,3)$ and of the
subalgebra $H=SO(1,3)$, respectively and $W_A^i(y)$ is the
infinitesimal $H$--compensator. Finally $L(y)$ is the coset
representative. Using eq.\eqn{mastformul} and denoting
 $\vec k_A =\{ \vec{\bf \Pi}, \vec{\bf N}, \vec {\bf J} \}$,
for the four translations we obtain the result:
\begin{eqnarray}
{\vec {\bf \Pi}}_0 &=&  \left( { {\rho}} \,t \right)
\,\frac{\partial}{\partial
\rho} \,
-\, \left( t\,x \right) \,\frac{\partial}{\partial x} \, - \,
{1\over2}( 1 + {{{\frac{1}{\rho^2}} }} + {t^2} + {w^2} + {x^2} )
\,\frac{\partial}{\partial t} \,  - \, \left( t\,w \right)
\,\frac{\partial}{\partial w} \,
 \nonumber\\
{\vec {\bf \Pi}}_1 &=&  \left( { {\rho}} \,w \right)
\,\frac{\partial}{\partial
\rho} \,
-\, \left( w\,x \right) \,\frac{\partial}{\partial x} \,  -\, \left(
t\,w
\right)
\,\frac{\partial}{\partial t} \, + \,
{1\over2}({1 + {{{\frac{1}{\rho^2}} }} - {t^2} - {w^2} + {x^2}})
\,\frac{\partial}{\partial w} \,
 \nonumber\\
{\vec {\bf \Pi}}_2 &=& -{ {\rho}} \,\frac{\partial}{\partial \rho}
\,
+ \,x
\,\frac{\partial}{\partial x} \, + \,t\,\frac{\partial}{\partial t}
\, + \,w
\,\frac{\partial}{\partial w} \,
 \nonumber\\
{\vec {\bf \Pi}}_3 &=& \left( { {\rho}} \,x \right)
\,\frac{\partial}{\partial
\rho} \, + \,
{1\over2}({1 + {\frac{1}{\rho^2} } - {t^2} + {w^2} - {x^2}})
\,\frac{\partial}{\partial x} \,  -\,\left( t\,x \right)
\,\frac{\partial}{\partial t} \,  -\,\left( w\,x \right)
\,\frac{\partial}{\partial w} \nonumber\\
\label{pkil}
\end{eqnarray}
while the three Lorentz boosts take the following form:
\begin{eqnarray}
{\vec {\bf N}}_1 &=&  -w\,
 \frac{\partial}{\partial t} \, - \,
t\, \frac{\partial}{\partial w} \,   \nonumber\\
{\vec {\bf N}}_2 &=& \left( { {\rho}} \,t \right)
\, \frac{\partial}{\partial \rho} \, - \,\left( t\,x \right)
\, \frac{\partial}{\partial x} \, + \,
{1\over2}({{1 - {{{\frac{1}{\rho^2}} }} - {t^2} - {w^2} - {x^2}}})
\, \frac{\partial}{\partial t} \, - \,\left( t\,w \right)
\, \frac{\partial}{\partial w} \,   \nonumber\\
{\vec {\bf N}}_3 &=&  -t
\, \frac{\partial}{\partial x} \, - \, x\, \frac{\partial}{\partial
t}
 \nonumber\\
 \label{nkil}
\end{eqnarray}
Finally for the three rotation generators we get:
\begin{eqnarray}
{\vec {\bf J}}_1 &=& \left( { {\rho}} \,w \right)
\, \frac{\partial}{\partial \rho} \, - \,\left( w\,x \right)
\, \frac{\partial}{\partial x} \, - \,\left( t\,w \right)
\, \frac{\partial}{\partial t} \, + \,
{1\over2}({{-1 + {{{\frac{1}{\rho^2}} }} - {t^2} - {w^2} + {x^2}}})
\, \frac{\partial}{\partial w} \,   \nonumber\\
{\vec {\bf J}}_2 &=&  -w
\, \frac{\partial}{\partial x} \, +  \,x
\, \frac{\partial}{\partial w} \,   \nonumber\\
{\vec {\bf J}}_3 &=& -{ {\rho}} \,x\, \frac{\partial}{\partial \rho}
\, + \,
{1\over2}({{1 - {{{\frac{1}{\rho^2}} }} + {t^2} - {w^2} + {x^2}}})
\, \frac{\partial}{\partial x} \, + \,t\,x
\, \frac{\partial}{\partial t} \, + \,w\,x
\, \frac{\partial}{\partial w} \nonumber\\
\label{jkil}
\end{eqnarray}
The corresponding compensating $SO(1,3)$ matrices are listed below.
\par
For the translations we get:
\begin{equation}
\begin{array}{cccccc}
 W \left ( \Pi_0 \right ) &=& - \left (\matrix{ 0 & 0 & w &
{\frac{1}{\rho}}  &
x \cr 0 & 0 & 0 & 0 & 0 \cr w & 0
& 0 & 0 & 0 \cr {\frac{1}{\rho}}  & 0 & 0 & 0 & 0 \cr x & 0 & 0 & 0
&
0 \cr  }
\right ) &
 W \left ( \Pi_1 \right ) &=&  \left (\matrix{ 0 & 0 &- t & 0 & 0
\cr
0 & 0 & 0
& 0 & 0 \cr -t & 0 & 0
&- {\frac{1}{\rho}}  &- x \cr 0 & 0 & {\frac{1}{\rho}}  & 0 & 0 \cr
0
& 0 & x &
0 & 0 \cr  } \right ) \\
\null & \null & \null & \null & \null & \null \\
 W \left ( \Pi_2 \right ) &=&  \left (\matrix{ 0 & 0 & 0 & 0 & 0 \cr
0 & 0 & 0
& 0 & 0 \cr 0 & 0 & 0
& 0 & 0 \cr 0 & 0 & 0 & 0 & 0 \cr 0 & 0 & 0 & 0 & 0 \cr  } \right )
&
 W \left ( \Pi_3 \right ) &=&  \left (\matrix{ 0 & 0 & 0 & 0 &- t
\cr
0 & 0 & 0
& 0 & 0 \cr 0 & 0 & 0
& 0 & w \cr 0 & 0 & 0 & 0 & {\frac{1}{\rho}}  \cr- t & 0 &- w &
-{\frac{1}{\rho}}  & 0 \cr  } \right )
\end{array}
\label{WKcomp}
\end{equation}
For the Lorentz boosts and the rotation generators the compensating
$SO(1,3)$ rotations are given below:
\begin{equation}
\begin{array}{cccccc}
 W \left ( N_1 \right ) &=&  \left (\matrix{ 0 & 0 & -1 & 0 & 0 \cr
0
& 0 & 0 &
0 & 0 \cr- 1 & 0 & 0
   & 0 & 0 \cr 0 & 0 & 0 & 0 & 0 \cr 0 & 0 & 0 & 0 & 0 \cr  } \right
)& W \left
( J_1 \right ) &=&
   \left (\matrix{ 0 & 0 & -t & 0 & 0 \cr 0 & 0 & 0 & 0 & 0 \cr- t &
0
& 0 &
-{\frac{1}{\rho}}  & -x \cr 0 & 0 &
   {\frac{1}{\rho}}  & 0 & 0 \cr 0 & 0 & x & 0 & 0 \cr  } \right
)\\
   \null & \null & \null & \null & \null & \null \\
 W \left ( N_2 \right ) &=&  -\left (\matrix{ 0 & 0 & w &
{\frac{1}{\rho}}  & x
\cr 0 & 0 & 0 & 0 & 0 \cr w & 0
   & 0 & 0 & 0 \cr {\frac{1}{\rho}}  & 0 & 0 & 0 & 0 \cr x & 0 & 0 &
0 & 0 \cr
} \right )& W \left ( J_2
   \right ) &=&  \left (\matrix{ 0 & 0 & 0 & 0 & 0 \cr 0 & 0 & 0 & 0
& 0 \cr 0
& 0 & 0 & 0 &- 1 \cr 0
   & 0 & 0 & 0 & 0 \cr 0 & 0 & 1 & 0 & 0 \cr  } \right )\\
  \null & \null & \null & \null & \null & \null \\
W \left ( N_3
  \right ) &=&  \left (\matrix{ 0 & 0 & 0 & 0 & -1 \cr 0 & 0 & 0 & 0
&
0 \cr 0 &
0 & 0 & 0 & 0 \cr 0
  & 0 & 0 & 0 & 0 \cr -1 & 0 & 0 & 0 & 0 \cr  } \right )&
  W \left ( J_3 \right ) &=&  \left (\matrix{ 0 & 0 & 0 & 0 & t \cr
0
& 0 & 0 &
0 & 0 \cr 0 & 0 & 0
  & 0 & -w \cr 0 & 0 & 0 & 0 & -{\frac{1}{\rho}}  \cr t & 0 & w &
{\frac{1}{\rho}}  & 0 \cr  } \right )
\\
\end{array}
\label{Wnjcomp}
\end{equation}
\subsection{Retrieving  broken conformal transformations}
When the four dimensional anti de Sitter group $SO(2,3)$ is
interpreted as the
conformal
group in three--dimensional space--time we are lead to split its
algebra as
follows:
\begin{eqnarray}
SO(2,3) &=& \left\{ P_\alpha \right\} \, \oplus \, \left\{
j_{\alpha\beta}
\right\}
\, \oplus  \, \left\{ K_\alpha \right\} \, \oplus \, {\cal D}
\nonumber\\
\left\{ P_\alpha \right\} & = & \mbox{translations} \quad ; \quad
\alpha = 0,1,2 \nonumber\\
\left\{ K_\alpha \right\} & = & \mbox{conformal boosts} \quad ;
\quad
\alpha = 0,1,3  \nonumber\\
\left\{ j_{\alpha\beta} \right\} & = & \mbox{Lorentz rotations}
\quad
; \quad
\alpha,\beta = 0,1,2 \nonumber\\
{\cal D} & = & \mbox{Dilatation}
\label{confdeco}
\end{eqnarray}
and consider its action the three--dimensional coset manifold
\begin{equation}
M^{Mink}_{1,2} = \frac{SO(2,3)}{D \times ISO(1,2)}
\label{minkio12}
\end{equation}
corresponding to $1+2$ Minkowski space. This leads to the standard
formulae for special conformal transformations on Minkowski
coordinates. What
should be noted, however is that the decomposition \eqn{confdeco} is
an
intrinsic
algebraic fact and it can be implemented in any case, also when
$SO(2,3)$ is realized as a group of isometries for four dimensional
anti de Sitter space. It just suffices to decide which is the group
$SO(1,2)$ that we want to consider as the three--dimensional Lorentz
group.  Our choice is the following. We identify the matrices of the
$SO(1,2)\subset SO(2,3)$ subalgebra with those of the following
form:
\begin{equation}
\left( \matrix{ 0 & 0 & 0 & 0 & 0 \cr
                  0   & \star & \star     &  0    & \star    \cr
                  0 & \star & \star & 0 &\star  \cr
                 0 & 0 & 0 & 0 & 0  \cr
                 0 & \star & \star & 0 &\star  \cr }
 \right)
 \label{lorentz12}
\end{equation}
and correspondingly, using the notations of eq.\eqn{vielgene} and
\eqn{nnnjjj},
we can easily identify
the translations and conformal boost generators as follows
\begin{equation}
\begin{array}{cc}
P_x = \Pi_3+J_3  \qquad & K_x = \frac{\Pi_3-J_3}{2}  \\
\null & \null  \\
P_t = -\Pi_0+N_2 \qquad
& K_t = -\frac{\Pi_0+N_2 }{2} \\
\null & \null \\
 P_w= \Pi_1-J_1  \qquad &
 K_w = \frac{ \Pi_1+J_1}{2}
\\
\end{array}
\label{pkalg}
\end{equation}
They form two separate three--dimensional abelian subalgebras of the
anti de Sitter algebra $SO(2,3)$:
\begin{equation}
\left[ P_\alpha \, , \, P_\beta \right] \, = \, \left[ K_\alpha \, ,
\, K_\beta
\right] \, \, = \, 0
\label{faczero}
\end{equation}
Using the explicit form of the Killing vectors \eqn{pkil},
\eqn{nkil}
and \eqn{jkil} we can write the transformations induced by the
operators
\begin{equation}
{\bf  P} \equiv p^x \, P_x \, +\,  p^t \, P_t \, + \, p^w \, P_w
\quad ; \quad
{\bf  K} \equiv k^x \, K_x \, +\,  k^t \, K_t \, + \, k^w \, K_w
\end{equation}
on the variables $\rho,x,t,w$. For the translations we immediately
find:
\begin{eqnarray}
\delta x&=& p^x \nonumber\\
\delta t &=&  p^t \nonumber\\
\delta w & = & p^w \nonumber\\
 \delta \rho & = & 0
\label{maldatrasl}
\end{eqnarray}
which is the result one would also obtain in ordinary Minkowski
space.
On the other hand for the conformal boosts one also immediately
obtains:
\begin{eqnarray}
\delta x&=& {k^t}\,t\,x - {k^w}\,w\,x +
  {\frac{{k^x}\,
      \left( \frac{1}{{\rho}^2} - {t^2} +
        {w^2} - {x^2} \right) }{2} } \nonumber\\
\delta t &=& -\left( {k^w}\,t\,w \right)  -
               {k^x}\,t\,x +
               {\frac{{k^t}\,
                   \left( \frac{1}{{\rho}^2} + {t^2} +
                     {w^2} + {x^2} \right) }{2} }\nonumber\\
\delta w & = & {k^t}\,t\,w - {k^x}\,x\,w +
                 {\frac{{k^w}\,
                     \left( \frac{1}{{\rho}^2} - {t^2} -
                       {w^2} + {x^2} \right) }{2}  } \nonumber\\
 \delta \rho & = & -
\rho\,\left( {k^t}\,t -
    {k^w}\,w - {k^x}\,x \right)
\label{maldaprim}
\end{eqnarray}
Naming the coordinates and the parameters as follows:
\begin{equation}
 y^\alpha =\left\{ x,t,w \right\} \quad ; \quad k^\alpha = \left\{
k^x
 , k^t , k^w \right\}
\label{nominal}
\end{equation}
and using the three dimensional
lorentzian metric $\eta^{\alpha \beta}$ $=\mbox{diag}(+,-,+)$ to
raise and lower  indices,  eq.s \eqn{maldaprim} can be rewritten as
follows:
\begin{eqnarray}
 \delta y^\alpha &=& - y^\alpha  \, y_\beta \, k^\beta \, + \,
 \frac{1}{2} \, k^\alpha \,\left( \frac{1}{\rho^2} \, + \, y_\beta
\,
y^\beta
 \right)\nonumber \\
 \delta \rho & = &
\rho\, \, y_\beta \, k^\beta
\label{maldasec}
\end{eqnarray}
Eq.\eqn{maldasec} exactly coincides with eq.(2.8) of the recent
paper
\cite{maldapasto} by Maldacena (it suffices to identify the
coordinate $\rho$
with the
field $U$. Indeed in \cite{maldapasto} the
transformations \eqn{maldasec} were interpreted as conformal
transformations on the $p$--brane world volume with respect to which
the microscopic  Born Infeld action is invariant. Their form is
similar
to that of canonical conformal transformations but not exactly
equal.
For this
reason they were named {\it broken conformal transformations}. Our
present
discussion
reveals their true meaning. They are indeed the transformations
generated by those generators of the $SO(2,3)$ group the act as
conformal boosts in the situation where $SO(2,3)$ is interpreted as
conformal group in $D=3$ dimensions. Their action however is not
calculated
on the three coordinates of the coset \eqn{minkio12}.
It is rather calculated on
the four coordinates of anti de Sitter space:
\begin{equation}
AdS_4 \, = \, \frac{SO(2,3)}{SO(1,3)}
\label{adscopro}
\end{equation}
which yields the result \eqn{maldasec}. The catch, however, is that
in the {\it solvable Lie algebra  parametrization} of $AdS_4$ the
Cartan coordinate $\rho$ is reinterpreted as a world--volume scalar
field.
So doing the isometry group of anti de Sitter space acts as a group
of field dependent conformal transformations for the world--volume
theory. We think that this explains the so far mysterious relation
between conformal symmetry of the world volume microscopic theory
and
the anti de Sitter symmetry of the near horizon geometry. We stress
that the clarifying item in this explanation is the choice of the
{\it solvable coordinates}.
\section{Conclusions and Perspective}
\label{conclu}
In this paper we have retrieved and discussed in a new perspective
the
class of BPS saturated
classical solutions of M--theory that are in one--to--one
correspondence with the old supersymmetric Freund--Rubin
compactifications
of $D=11$ supergravity and reduce to them on the horizon.
We have shown the relation between the anti de Sitter symmetry
of the Freund--Rubin compactification and the superconformal
symmetry of the world--volume theory.
\par
In particular, our discussion suggests that there should exist
microscopic world--volume theories where the superisometry group
${\cal SC}^{G/H}$  (see
\eqn{scGH}) of the  supergravity theory is realized as
a global symmetry group. Furthermore, one should be able to
reconstruct
the massless states of supergravity
that belong to specific representations of ${\cal SC}^{G/H}$, as
suitable
tensor products of `singleton' representations of ${\cal SC}^{G/H}$.
 This mechanism
has already been verified \cite{fronsd} for ordinary branes, where
the
superconformal group $\cal SC$ is a simple supergroup $Osp(N\vert
4)$.
In our case the novelty is the existence of the residual symmetry
group $G'$
so that the group theoretical construction of the massless states
should agree at the level of both factors. This suggests a
microscopic
world volume theory with suitable matter multiplets. A search for
these
conjectured world--volume theories is postponed to the future.
\par
In addition, we should stress that the construction of generalized
M--branes,
interpolating between Kaluza--Klein vacua at the horizon and flat
manifolds
at spatial infinity, does not exhaust all the possibilities.
Specifically, the ansatz  \eqn{Ansatz1} can be further generalized by
replacing
the {\it angular} part  of the metric $ds^2_{G/H}$ with a general
 Einstein metric that admits no continuous isometry (as already
noted
 in \cite{duffetal}). Typically this is achieved by
orbifoldizing the coset manifold $G/H$ with respect to the action of
some
discrete subgroup $\Gamma \subset G$.
Quite likely such a procedure
produces models similar to those already considered from a
microscopic world-volume point of view in \cite{lowersusy}.
\par
It is now a challenging problem to retrieve a description of
these string solitons in string language, in particular in terms
of $D$--branes.
\appendix
\section{$AdS_{p+2}$ as a solvable group manifold}
\label{mario}
The aim of the present appendix is to show how anti de Sitter space
in $d+1$ dimensions, {\it non--compact, lorentzian coset manifold}
\begin{equation}
 AdS_{d+1}  \, \equiv \, \frac{SO(2,d)}{SO(1,d)}
 \label{dp1ads}
\end{equation}
can be described as a {\it solvable group manifold}.
As anticipated in the main text,
our analysis extends to a pseudo--riemanian case  the classical
treatment of riemanian non compact homogeneous manifolds
\cite{helgason}
we have already extensively utilized to discuss the supergravity
scalar
sectors \cite{lucianoi},\cite{noi1},\cite{noi2}, \cite{noi3}.
\par
Specifically, in this appendix we describe the structure of the
solvable
Lie algebra defined by the decomposition:
\begin{equation}
SO(2,d)\, =\, SO(1,d)\oplus Solv
\label{iwads}
\end{equation}
Our result is that the structure of $Solv$ can be described as
follows:
\begin{equation}
Solv\, =\, {\cal C}_K\oplus (\sum_{\alpha\in
\Delta^+}E_{\alpha})\cap
SO(2,d)
\label{solv}
\end{equation}
where ${\cal C}_K$ denotes the one--dimensional
space consisting of the unique non--compact Cartan generator of
$SO(2,d)$ which is not contained in $SO(1,d)$ and which
therefore enters the quotient $SO(2,d)/SO(1,d)$.
On the other hand the space $\Delta^+$
consists of all the roots of $SO(2,d)$ that have  a strictly
positive
value on the Cartan generator in ${\cal C}_K$.
The intersection symbol in eq. \eqn{solv} is used because, in
general, the {\it shift} operators $E_{\alpha}\,
\alpha\in \Delta^+$   do not belong to the $SO(2,d)$ real
form of the $SO(2+d)$ complex algebra. However there are  suitable
linear
combinations
of these operators which do belong to such a real form.
Hence {\it shift} operators $E_{\alpha}\, \alpha\in \Delta^+$
will enter the structure of $Solv$ defined by eq. \eqn{solv}
only through   the appropriate linear combinations.
\par
 In what follows we first give the explicit representation of the
$Solv$ generators in terms of $(d+2)\times (d+2)$  matrices leaving
the metric  $\eta=\mbox{diag}\left\{ -, - , + , \dots , + \right\}$
invariant.  Then a derivation of eq. \eqn{iwads}
will be illustrated in more abstract terms using the Iwasawa
decomposition.
\par
The root system of $SO(2,d)$ can be expressed, with respect to an
orthonormal basis
\begin{equation}
 \{\epsilon_i\}_{i=1,\dots, r}
\end{equation}
of $\IR^r$, $r=rankSO(2,d)$, in the following way:
\begin{equation}
\Phi\, =\, \cases{\pm\epsilon_i\pm\epsilon_j\,\,1\le i<j\le r &
d+2=2r\cr
\pm\epsilon_i\pm\epsilon_j\,\,1\le i<j\le r\,,\,\pm
\epsilon_i\,\,i=1,...,r
& d+2=2r+1 }
\label{rotsys}
\end{equation}
The non compact Cartan generators of $SO(2,d)$ are
$\{H_{\epsilon_1},H_{\epsilon_2}\}$. Choosing the $SO(1,d)$
subalgebra
of $SO(2,d)$ that admits $H_{\epsilon_1}$ as the non--compact
Cartan generator, the space ${\cal C}_K$ in eq. \eqn{solv} will
consist
of $H_{\epsilon_2}$ only. Then the roots in $\Delta^+$ will be:
\begin{equation}
\Delta^+\, =\, \cases{\epsilon_2\pm\epsilon_1\,,
\,\epsilon_2\pm\epsilon_i\,\,i=3,...,r & 2+d=2r \cr
\epsilon_2 \,,\,\epsilon_2\pm\epsilon_1\,,
\,\epsilon_2\pm\epsilon_i\,\,i=3,...,r & 2+d=2r+1}
\label{delta+}
\end{equation}
In order to construct the $(d+2)$--dimensional
matrix representation of the $SO(2,d)$ generators and
therefore of the operators in $Solv$, we start by defining
the non--compact Cartan
generators $\{H_{\epsilon_1},H_{\epsilon_2}\}$ and the compact ones
$\{{\rm i}H_{\epsilon_i} \,\,\,i=3,...,r\}$ as $(d+2)\times (d+2)$
matrices
 whose non vanishing entries are given by:
\begin{eqnarray}
(H_{\epsilon_1})_{1,3}\,&=&\, (H_{\epsilon_1})_{3,1}\, =\, 1
\,;\,(H_{\epsilon_2})_{2,4}\, =\,(H_{\epsilon_2})_{4,2}\, =\,
1\nonumber\\
({\rm i}H_{\epsilon_{k+2}})_{2(k+1)+1,2(k+1)+2}\, &=&\,-({\rm
i}H_{\epsilon_{k+2}})_{2(k+1)+2,2(k+1)+1}\, =\, 1\,\,\,\,k=1,\dots
,r-2
\label{cartan}
\end{eqnarray}
The {\it shift} operators are  represented by eigenmatrices of
the adjoint action of the Cartan operators in eqs. \eqn{cartan}.
Adopting  a suitable convention on the normalization of the
{\it shift} operators, it follows that the matrices
representing
$E_{\pm\epsilon_2\pm\epsilon_1}\,,\,E_{\epsilon_2}\,,\,E_{\epsilon_1}$ 

 are
in the $SO(2,d)$ real form. In particular, the
$E_{\epsilon_2\pm\epsilon_1}$ matrices are characterized by non
vanishing entries only in the upper
$5\times 5$ diagonal blocks which coincide respectively with the
$SO(2,3)$ representation (${\cal T}_{\mp}$) of the same operators
given in
eq. \eqn{tpm}. Moreover in the $d$--odd case the matrix realization
of $E_{\epsilon_2}$ is defined by the following non--zero entries:
\begin{eqnarray}
(E_{\epsilon_2})_{2,d+2}\,&=&\,(E_{\epsilon_2})_{d+2,2}\, =\,
\frac{1}{\sqrt{2}}\nonumber\\
(E_{\epsilon_2})_{4,d+2}\,&=&\,-(E_{\epsilon_2})_{d+2,4}\, =\,
\frac{1}{\sqrt{2}}
\label{epsilon12}
\end{eqnarray}
The operators $E_{\epsilon_2\pm\epsilon_i}\,\,i=3,...,r $ are
represented by
complex matrices whose real and imaginary parts
separately belong to $SO(2,d)$. Moreover we can normalize each
matrix
so  that $E_{\epsilon_2-\epsilon_i}=(E_{\epsilon_2+\epsilon_i})^*$.
Thus
the generators $E_{\epsilon_2\pm\epsilon_i}\,\,i=3,...,r $ will
enter
the formula \eqn{solv} only through the following combinations which
single
out  their real and imaginary parts:
\begin{eqnarray}
X_1&=&E_{\epsilon_2+\epsilon_3}
+E_{\epsilon_2-\epsilon_3}\,,\,X_2=
-{\rm i}(E_{\epsilon_2+\epsilon_3}-
E_{\epsilon_2-\epsilon_3})\dots
\nonumber\\
X_{2r-5}&=&E_{\epsilon_2+\epsilon_{r-1}}+E_{\epsilon_2-\epsilon_{r-1}} 
\, ,\,X_{2r-4}=-{\rm
i}(E_{\epsilon_2+\epsilon_r}-E_{\epsilon_2-\epsilon_r})
\label{Xi}
\end{eqnarray}
The corresponding matrix representation in the $d+2=2r$ case
is characterized by the following non--zero entries:
\begin{eqnarray}
(X_k)_{2,4+k}\,&=&\,(X_k)_{4+k, 2}\,=\,1\nonumber\\
(X_k)_{4,4+k}\,&=&\,-(X_k)_{4+k, 4}\,=\,1\,\,\,k=1,\dots,2r-4
\label{Ximatrix}
\end{eqnarray}
The matrices $X_k$ defined above in the $2+d=2r$ case are $d-2$.
In the case $2+d=2r+1$, the set of $d-2$ matrices $X_k$ is completed
by defining
$X_{d-2}/\sqrt{2}=E_{\epsilon_2}$, whose matrix representation is
given in eq. \eqn{epsilon12}.
The usefulness of this notation will become apparent once we
interpret
the parameters of the solvable algebra $Solv$ as the coordinates on
the world wolume of a $(p=d-1)$--brane: $(t,w,x^1,\dots,x^{p-1})$.
Indeed the generators  of $Solv$, according to the structure
described in eq.\eqn{solv}, can now be written in the form:
\begin{equation}
Solv\,=\, \{H_{\epsilon_2}, {\cal T}_{-}=E_{\epsilon_2+\epsilon_1},
{\cal T}_{+}=E_{\epsilon_2-\epsilon_1}, X_1,\dots,X_{p-1}\}
\label{solvspec}
\end{equation}
Let us define the coset representative of the $AdS$
coset space \eqn{dp1ads}
as a solvable group element generated by a combination
of the $Solv$ generators as:
\begin{eqnarray}
\IL(a,t,w,x^1,\dots,x^{p-1})\,&=&\,\tau (t,w,x^1,\dots,x^{p-1})
\cdot S(a)\nonumber\\
\tau (t,w,x^1,\dots,x^{p-1})\,&=&\,exp\left[(t-w){\cal T}_{-}+
(t+w){\cal T}_{+}+x^1X_1+\dots +x^{p-1}X_{p-1}\right]\nonumber\\
S(a)\,&=&\,exp\left[-a H_{\epsilon_2}\right]
\label{coset}
\end{eqnarray}
This is the generalization to the generic $d+1$ dimensional case
of eq. \eqn{cosettus} of the main text corresponding to the $d=3$
case. Computing the left invariant 1--form one finds:
{\scriptsize
\begin{equation}
\Omega\,=\,\IL^{-1}d\IL\,=\,\left(\begin{tabular}{cc|ccccccc}
0 & $-dt \, e^a$  & 0 & $dt \, e^a$ &  0 & 0  & \dots & 0 & 0 \\
 $dt \, e^a$ & 0  & $dw \, e^a$ & $-da$ & $dx^1 \, e^a$ & $dx^2 \,
 e^a$  & \dots &  $dx^{p-2} \, e^a$ &
$dx^{p-1} \, e^a$  \\ \hline
 0 & $dw \, e^a$ & 0 & $-dw \, e^a$ &  0 & 0  & \dots &  0 & 0  \\
$dt \, e^a$  & $-da$  & $dw \, e^a$ & 0 & $dx^1 \, e^a$ & $dx^2 \,
e^a$  &
\dots &  $dx^{p-2} \, e^a$ &
$dx^{p-1} \, e^a$  \\
 0  &  $dx^1 \, e^a$ & 0  & $-dx^1 \, e^a$ & 0 & 0  & \dots & 0 & 0
\\ 0
 &  $dx^2 \, e^a$ & 0  & $-dx^2 \, e^a$ & 0 & 0  &
\dots & 0 & 0 \\ $\vdots $ &$\vdots $  & $\vdots $  & $\vdots $  &
$\vdots $  &
$\vdots $  &$ \ddots$ &  $\vdots $ &$\vdots $ \\
 0  & $dx^{p-2} \, e^a$ & 0  & $-dx^{p-2} \, e^a$ & 0 & 0 & \dots &
0
& 0\\
  0  &  $dx^{p-1} \, e^a$ & 0  & $-dx^{p-1} \, e^a$ & 0 & 0 & \dots
&
0 & 0
\end{tabular}
\right)
\label{1formgen}
\end{equation}}
which is also the straightforward generalization
of the left invariant 1--form computed in eq. \eqn{leftinv} for the
$SO(2,3)$ case. The role of the transverse coordinate
$x_{\perp}$ now is played by the $p-1$ parameters $x^i$.
As in the $p=2$ case,
the parameters $t-w$ and $t+w$ are the light--cone coordinates on
the world volume of the p--brane.
Indeed extending the same procedure previously followed
for the $p=2$ case, it is straightforward to compute the vielbein
in the  general case of $AdS^{p+2}$ we are considering:
\begin{eqnarray}
V^0\, &=& \,-\frac{1}{2}{\rm Tr}(\Pi_0\Omega)\nonumber\\
V^k\, &=& \,\frac{1}{2}{\rm Tr}(\Pi_k\Omega)\,\,\,k=1,\dots,p+1
\end{eqnarray}
where $\{\Pi_0,\Pi_k\}$ is the basis of matrices defined by the
orthogonal decomposition of $SO(2,p+1)$ with respect to $SO(1,p+1)$:
\begin{eqnarray}
\Pi_0\, &=&\, M^{1,2}\nonumber\\
\Pi_k\, &=&\, M^{1,2+k}\,\,\,k=1,\dots ,p+1
\end{eqnarray}
$M^{A,B}$ being the orthogonal basis of $SO(2,d)$ generators.
The expression for the vielbein is:
\begin{equation}
\left(\matrix{V^0\cr V^1\cr V^2 \cr V^3\cr \vdots \cr
V^{p+2}}\right)\, =\,
\left(\matrix{dt \, e^a \cr dw \, e^a\cr -da \cr dx^1 \, e^a\cr
\vdots \cr
dx^{p-1} \, e^a }\right)
\label{vielbeingen}
\end{equation}
Setting $ \rho\equiv e^a$, from the vielbein basis we can compute
the
metric:
\begin{eqnarray}
ds^2\, &=&\,-V^0\otimes V^0+\sum_{k=1}^{p+1} V^k\otimes V^k
\nonumber
\\
&=&\, \rho^2 (-dt^2+dw^2+d\vec{x}\cdot d\vec{x})+\rho^{-2}d\rho^2
\label{bertottigen}
\end{eqnarray}
which is the $AdS^{p+2}$ metric in  Bertotti--Robinson form.
It is worthwhile noticing that it is possible to characterize
in an intrinsic geometrical way the coordinates on the world sheet
of the p--brane as the parameters of the {\it maximal abelian ideal}
${\cal A}$ of the solvable Lie algebra generating the
$AdS^{p+2}$ space. Indeed it is straightforward to check that:
\begin{equation}
{\cal A}\, =\,\{{\cal T}_+,{\cal T}_-, X_1,\dots,X_{p-1}\}\subset
Solv
\end{equation}
is the maximal abelian ideal of $Solv$.
\subsection{The solvable algebra and Iwasawa decomposition}
To conclude  we  resume our discussion from
the start and we give a derivation of eq. \eqn{iwads} which allows
the solvable Lie algebra description of $AdS^{p+2}$.
Let us first consider the Iwasawa decomposition of $SO(2,d)$ and
of $SO(1,d)$ separately. At this point
it is useful  to recall
the main features of the Iwasawa decomposition
of a non--compact semisimple Lie algebra.
Any non--compact real form $\IG_o$ of a complex semisimple Lie
algebra $\IG$ can be represented, according to the Iwasawa
decomposition,
as the direct sum of its maximal compact subalgebra $\IH$ and a
solvable Lie algebra $Solv$:
\begin{equation}
\IG_o\, =\, \IH\oplus Solv
\end{equation}
The structure of $Solv$ is the  following:
\begin{eqnarray}
Solv\, &=&\, {\cal C}_{nc}\oplus {\cal N}il\nonumber\\
{\cal N}il\, &=&\,(\sum_{\alpha\in \Delta^+_{o}}E_{\alpha})\cap
\IG_o
\end{eqnarray}
where ${\cal C}_{nc}$ is the subspace of all the non--compact
Cartan generators
and the remaining nilpotent part ${\cal N}il$ of $Solv$ is generated
by all the {\it shift} generators of $\IG$ associated with roots
which
are positive with respect to ${\cal C}_{nc}$ and do not vanish
identically on it.
Moreover these {\it shift} generators have to be suitably combined
with each other in order to obtain nilpotent operators in the real
form
$\IG_o$.
The Iwasawa decomposition for $SO(2,d)$ and $SO(1,d)$ reads as
follows
\begin{eqnarray}
SO(2,d)\, &=&\, SO(2)\oplus SO(d) \oplus Solv_{SO(2,d)}\nonumber \\
 Solv_{SO(2,d)}\, &=&\, {\cal C}_{SO(2,d)}\oplus {\cal
N}il_{SO(2,d)}
 \nonumber \\
SO(1,d)\, &=&\, SO(d) \oplus Solv_{SO(1,d)}\nonumber \\
 Solv_{SO(1,d)}\, &=&\, {\cal C}_{SO(1,d)}\oplus {\cal
N}il_{SO(1,d)}
\label{iwasawa12}
\end{eqnarray}
where ${\cal C}_{SO(2,d)}=\{H_{\epsilon_1},H_{\epsilon_2}\}$ is
the space spanned by the non--compact Cartan generators of $SO(2,d)$
and
${\cal C}_{SO(1,d)}=\{H_{\epsilon_1}\}$ consists of the unique
non--compact Cartan generator belonging to the chosen $SO(1,d)$
subgroup of
$SO(2,d)$.
In order to simplify the notation let us define a set of $d-2$
nilpotent generators $Y_i$ in the same way as for the $X_i$
generators:
\begin{eqnarray}
2+d\, &=&\, 2r\nonumber \\
Y_1\, &=&\,E_{\epsilon_1+\epsilon_3} +E_{\epsilon_1-\epsilon_3}\,,
\,Y_2=-{\rm
i}(E_{\epsilon_1+\epsilon_3}-E_{\epsilon_1-\epsilon_3})\dots
\nonumber \\
& &
Y_{2r-5}=E_{\epsilon_1+\epsilon_{r-1}}+E_{\epsilon_1-\epsilon_{r-1}}\,  ,
\,Y_{2r-4}=-{\rm
i}(E_{\epsilon_1+\epsilon_r}-E_{\epsilon_1-\epsilon_r})
\label{Yi}
\end{eqnarray}
For $2+d=2r+1$ let us define $Y_{d-2}/\sqrt{2}=E_{\epsilon_1}$.
The structure of $Solv_{SO(2,d)}$ and of $Solv_{SO(1,d)}$ can be
described as:
\begin{eqnarray}
Solv_{SO(2,d)}\, &=&\, \{H_{\epsilon_1},H_{\epsilon_2}\}\oplus
{\cal N}il_{SO(2
,d)}\nonumber\\
{\cal N}il_{SO(2,d)}\, &=&\,\{ E_{\epsilon_1\pm \epsilon_2}, X_i,
Y_i,\,i=1,\dots, d-2\}\nonumber\\
Solv_{SO(1,d)}\, &=&\, \{H_{\epsilon_1}\}\oplus {\cal N}il_{SO(1,d)}
\nonumber\\
{\cal N}il_{SO(1,d)}\, &=&\,\{ (E_{\epsilon_1+ \epsilon_2}+
E_{\epsilon_1- \epsilon_2}),Y_i,\,i=1,\dots, d-2\}
\end{eqnarray}
The operators $E_{\epsilon_1+ \epsilon_2}$ and
$E_{\epsilon_1- \epsilon_2}=(E_{-\epsilon_1 +\epsilon_2})^t$,
in our matrix representation, enter $SO(1,d)$ only through their
sum.
This can be seen directly from their matrix representation and from
the fact that our choice   $SO(1,d) \subset SO(2,d)$ corresponds to
the
group of $SO(2,d)$ matrices having zero entries along the second row
and the second column. Indeed the matrix representation of
$E_{\epsilon_1- \epsilon_2}+E_{\epsilon_1+ \epsilon_2}$ has the
following non--zero entries:
\begin{eqnarray}
(E_{\epsilon_1- \epsilon_2}+E_{\epsilon_1+ \epsilon_2})_{1,4}\,&=&
\,(E_{\epsilon_1- \epsilon_2}+E_{\epsilon_1+ \epsilon_2})_{4,1}\,
=\,
\frac{1}{2}\nonumber \\
(E_{\epsilon_1- \epsilon_2}+E_{\epsilon_1+ \epsilon_2})_{3,4}\,&=&
\,-(E_{\epsilon_1- \epsilon_2}+E_{\epsilon_1+ \epsilon_2})_{4,3}\,
=\, \frac{1}{2}
\label{combination}
\end{eqnarray}
It is immediate to check that  also the combination
$E_{-(\epsilon_1+ \epsilon_2)}+E_{-\epsilon_1+ \epsilon_2}$,
represented by the transpose of the matrix in eq.\eqn{combination},
belongs to $SO(1,d)$.
In the basis defined by the Iwasawa decompositions \eqn{iwasawa12},
we can write:
\begin{eqnarray}
SO(2,d)\, &=&\, SO(1,d)\oplus SO(2)\oplus \{Solv_{SO(2,d)}/Solv_
{SO(1,d)}\}\nonumber\\
SO(2)\, &=&\,\{g\} \,=\,\{ E_{\epsilon_1+ \epsilon_2}
-E_{\epsilon_1- \epsilon_2}-E_{-(\epsilon_1+ \epsilon_2)}
+E_{-\epsilon_1+ \epsilon_2}\}\,=\,K_0\nonumber \\
Solv_{SO(2,d)}/Solv_{SO(1,d)} \, &=&\,\{ H_{\epsilon_2},
E_{\epsilon_1+ \epsilon_2}-E_{\epsilon_1- \epsilon_2}, X_i, \,
i=1,\dots ,d-2\}
\label{iwabasis}
\end{eqnarray}
It is straightforward to verify that
we can perform a transformation on the basis defined in eqs.
\eqn{iwabasis}
by means of which the generator $g$ of $SO(2)$ and the generator
$E_{ \epsilon _1+ \epsilon _2}-E_{\epsilon _1- \epsilon _2}$ in
$Solv_{SO(2,d)}/Solv_{SO(1,d)}$
are mixed with the two operators $\{E_{\epsilon_1+ \epsilon_2}
+E_{\epsilon_1- \epsilon_2},E_{-(\epsilon_1+ \epsilon_2)}
+E_{-\epsilon_1+ \epsilon_2}\} $ in $SO(1,d)$ to obtain two
independent combinations represented by the operators
$E_{\epsilon_1+ \epsilon_2}$ and $E_{-\epsilon_1+ \epsilon_2}$.
The latter, together with the remaining generators in
$Solv_{SO(2,d)}/Solv_{SO(1,d)}$ define the following solvable Lie
algebra:
\begin{equation}
Solv\,=\,\{ H_{\epsilon_2},E_{\epsilon_1+ \epsilon_2},
E_{-\epsilon_1+ \epsilon_2}  X_i, \, i=1,\dots ,d-2\}
\end{equation}
This result  coincides with the one in eq. \eqn{solvspec} for
$d=p+1$.
Therefore this new basis realizes
the decomposition \eqn{iwads}.


\begin{thebibliography}{77}
\bibitem{secrevol1} E. Witten, ``String Theory Dynamics in various
dimensions", Nucl. Phys. {\bf B443} (1995) 85;
C.M. Hull, P.K. Townsend, ``Unity of String Dualities", Nucl. Phys.
{\bf B438}
(1995) 109
\bibitem{secrevol2} P. Horava, E. Witten,
``Heterotic and Type I String Dynamics from Eleven Dimensions",
Nucl. Phys. {\bf B460} (1996) 506
\bibitem{sugra11a} E. Cremmer, B. Julia, J. Scherk
``Supergravity theory in eleven dimensions"
{\it Phys. Lett.} {\bf B76} (1978) 409
\bibitem{sugra11b} R. D'Auria, P. Fr\'e ,``Geometric
supergravity in D=11 and its hidden supergroup",
{\it Nucl.Phys.} {\bf B201} (1982) 101
\bibitem{mbrastelle} The literature on this topic is quite
extended. As a general review, see the lecture notes:
\par
K. Stelle, ``Lectures on Supergravity p--Branes'',
Lectures presented at 1996 ICTP Summer School, Trieste,
hep-th/9701088
\bibitem{mbratownsend} For a recent comprehensive updating on
M--brane solutions see also
\par
P. K. Townsend, ``M--Theory from its Superalgebra'',
Talk given at the NATO Advanced Study
Institute on Strings, Branes and Dualities, Cargese, France,
26 May - 14 June 1997, hep-th/9712004
\bibitem{polchidbra}
J. Polchinski, ``TASI Lectures on D--Branes'',
hep-th/9611050; J. Polchinski, S. Chaudhuri and C. V.Johnson,
``Notes on D-Branes'', hep-th/9602052
\bibitem{bstastef} P. Di Vecchia, M. Frau, I. Pesando, S. Sciuto,
 A. Lerda and
R. Russo, ``Classical P--Branes from Boundary States'',
  {\it Nucl. Phys.} {\bf B507} , 259 (1997), hep-th/9707068
\bibitem{bstaieng}
M. Bertolini, R. Iengo and C. A. Scrucca,
``Electric and Magnetic Interaction of Dyonic D--Branes and Odd
Spin Structures'', hep-th/9801110
\bibitem{bstapaol}
 M. Bill\`o , P. Di Vecchia , M. Frau , A. Lerda, I. Pesando,  R.
Russo and S. Sciuto , ``Microscopic String Analysis of the D0--Brane
System and Dual RR States'', hep-th/9802088
\bibitem{freundrub}
P.G.O. Freund and M.A. Rubin, ``Dynamics of Dimensional Reduction'',
 {\it Phys. Lett} {\bf 97B} (1980) 233
\bibitem{KKduff} M.J. Duff, B.E.W. Nilsson and C.N. Pope,
``Kaluza Klein Supergravity'', {\it Phys. Rep.}
{\bf 130} (1986) 1
\bibitem{KKenglert} F. Englert, ``Spontaneous Compactification of
11--Dimensional Supergravity'',
 {\it Phys. Lett} {\bf 119B} (1982) 339
\bibitem{mpqr} L. Castellani, R. D'Auria and P. Fr\`e,
``$SU(3)\times SU(2)\times U(1)$ from D=11 Supergravity'', {\it
Nucl. Phys. } {\bf 239} (1984) 60
\bibitem{KKwarncastel} L. Castellani, L. J. Romans and N. P. Warner,
``A Classification of Compactifying Solutions of D=11
Supergravity'',
{\it Nucl. Phys. } {\bf B241} (1984) 429
\bibitem{duffetal} M.J.Duff, H. L\"u, C.N. Pope and E. Sezgin,
``Supermembranes with fewer supersymmetries", Phys. Lett. {\bf B371}
(1996) 206, hep-th/9511162.
\bibitem{renatoine} P. Claus, R. Kallosh, and A. Van Proeyen,
`` M5--Brane and Superconformal (0,2) Tensor Multiplet in
Six-Dimensions'', hep-th/9711161;
P. Claus, R. Kallosh, J. Kumar, P. Townsend and A. Van Proeyen,
`` Conformal Theory of M2, D3, M5 and D1+D5 Branes'', hep-th/9801206
\bibitem{maldapasto}J. Maldacena, ``The Large N Limit of
Superconformal
Field Theories and Supergravity'', hep-th/9711200
\bibitem{doubling}
G. Gibbons and P. Townsend, ``Vacuum Interpolation in Supergravity
via Super p-Branes'',{\it Phys. Rev. Lett.} {\bf 71} (1993) 5223;
G. Gibbons, {\it Nucl. Phys.} {\bf B207} (1982) 337;
R. Kallosh and A. Peet, {\it Phys. Rev.} {\bf B46} (1992) 5223,
hep-th/9209116;
S. Ferrara, G. Gibbons and R. Kallosh, {\it Nucl. Phys.} {\bf B500}
(1997) 75, hep-th/9702103;
A. Chamseddine, S. Ferrara, G.W. Gibbons and R. Kallosh,
``Enhancement of
Supersymmetry Near 5--D Black Hole Horizon'', {\it Phys. Rev.}
{\bf D55} 3647 (1997), hep-th/9610155
\bibitem{pietricmspec} R. D'Auria and P. Fr\'e,
``On the Fermion Mass Spectrum of Kaluza--Klein Supergravity'',
{\it Ann. of Phys.} {\bf 157} (1984) 1;
R. D'Auria and P. Fr\'e, ``Universal Bose--Fermi Mass Relations and
Harmonic Analysis on Coset Manifolds with Killing spinors,
{\it Ann. of Phys.} {\bf 162} (1985) 372.
\bibitem{bertotrobin} B. Bertotti, {\it Phys. Rev} {\bf 116} (1959)
1331;
I. Robinson, {\it Bull. Acad. Polon.} {\bf 7} (1959) 351
\bibitem{castdauriafre} L. Castellani,  R. D'Auria and  P. Fr\'e,
``Supergravity and Superstrings --A Geometric Perspective'',
vol 1,2,3, World Scientific (1991)
\bibitem{noi1}
L. Andrianopoli, R. D'Auria, S. Ferrara, P. Fr\'e,
and M. Trigiante {\it Nucl. Phys.} {\bf B496} (1997) 617,
hep-th/9611014
\bibitem{noi2}
L. Andrianopoli, R. D'Auria, S. Ferrara, P. Fr\'e,
and M. Trigiante {\it Nucl. Phys.} {\bf B493} (1997) 249,
hep-th/9612202
\bibitem{noi3}
L. Andrianopoli, R. D'Auria, S. Ferrara, P. Fr\'e,
and M. Trigiante {\it Nucl.Phys.} {\bf B509} (1998) 463,
hep-th/9707087
\bibitem{lucianoi}
P. Fr\'e, L. Girardello, I. Pesando
and M. Trigiante {\it Nucl.Phys.} {\bf B493} (1997) 231,
hep-th/9607032
\bibitem{alex}
D.V. Alekseevskii, Math. USSR Izvestija, Vol. {\bf 9} (1975), No.2
\bibitem{helgason}
S. Helgason, `` {\it Differential Geometry and Symmetric Spaces}'',
New York: Academic Press (1962)
3
\bibitem{holow}E. Witten, ``Anti de Sitter Space and Holography'',
hep-th/9802150
\bibitem{nairdaemi}
V. P. Nair and S. Randjbar-Daemi, ``On brane solutions
in M(atrix) theory'', hep-th/9802187
\bibitem{lowersusy}
 N. Itzhaki, J.M. Maldacena, J. Sonnenschein and S. Yankielowicz,
``Supergravity and the Large N Limit of Theories with Sixteen
Supercharges'',
hep-th/9802042; M. Berkooz, ``A Supergravity Dual of a (1,0) Field
Theory in
Six Dimensions'', hep-th/9802195; S. Kachru and E. Silverstein, ``4D
Conformal
Theories and Strings on Orbifolds'', hep-th/9802183;
A. Lawrence and C. Vafa, ``On conformal Field Theories in Four
Dimensions''
hep-th/9803015
\bibitem{fronsd} S. Ferrara and C. Fronsdal, ``Gauge fields as
composite
boundary excitations'',
hep-th/9802126
\bibitem{kalkum}R. Kallosh,  J. Kumar and A. Rajaraman,
``Special Conformal Symmetry of World Volume Actions'',
hep-th/9712073
\bibitem{zaffa} S. Ferrara, C. Fronsdal and A. Zaffaroni,
``On N=8 Supergravity on $AdS_5$ and N=4 Superconformal Yang-Mills
theory'',
hep-th/9802203,
S. Ferrara, A. Zaffaroni, in preparation.
\end{thebibliography}
\end{document}